\documentclass[pre,aps]{revtex4}

\usepackage{times}
\usepackage{amsmath}
\usepackage{graphicx}
\usepackage{amssymb}
\usepackage{color}

\newcommand{\wb}{\omega_{\mathrm{b}}}
\newcommand{\Tb}{T_{\mathrm{b}}}

\newcommand{\ii}{\mathrm{i}}

\newcommand{\F}{\mathcal{F}}

\newcommand{\cP}{\ensuremath{\mathcal{P}}}
\newcommand{\cT}{\ensuremath{\mathcal{T}}}

\begin{document}

\title{$\cP\cT$-Symmetric Dimer of Coupled Nonlinear Oscillators}

\author{Jes\'us Cuevas}
\affiliation{Nonlinear Physics Group, Departamento de F\'{\i}sica Aplicada I, Universidad de Sevilla. Escuela Polit{\'e}cnica Superior, C/ Virgen de \'Africa, 7, 41011-Sevilla, Spain}

\author{Panayotis\ G.\ Kevrekidis}
\affiliation{Department of of Mathematics and Statistics, University of
Massachusetts, Amherst, MA 01003-9305, USA}

\author{Avadh Saxena}
\affiliation{Center for Nonlinear Studies and Theoretical Division, Los Alamos National Laboratory, Los Alamos, New Mexico 87545, USA}

\author{Avinash Khare}
\affiliation{Indian Institute of Science Education and Research (IISER), Pune 411008, India}

\begin{abstract}
We provide a systematic analysis of a prototypical
nonlinear oscillator system respecting $\cP\cT$-symmetry i.e., one
of them has gain and the other an equal and opposite amount of loss.
Starting from the linear limit of the system, we extend considerations
to the nonlinear case for both soft and hard cubic nonlinearities
identifying symmetric and anti-symmetric breather
solutions, as well as symmetry breaking variants thereof.
We propose a reduction of the system to a Schr{\"o}dinger type
 $\cP\cT$-symmetric dimer, whose detailed earlier understanding
can explain many of the phenomena observed herein, including
the $\cP\cT$ phase transition. Nevertheless,
there are also significant parametric as well as phenomenological
potential differences between the two models and we discuss where these
arise and where they are most pronounced. Finally, we also provide
examples of the evolution dynamics of the different states in their
regimes of instability.
\end{abstract}
\maketitle

\section{Introduction}

The topic of Parity-Time ($\mathcal{PT}$) symmetry and its relevance
to physical applications on the one hand, as well as its mathematical
structure on the other have drawn considerable attention both from the
physics and the mathematics communities. Originally, this theme
was proposed by C. Bender and co-workers as an additional
possibility for operators associated with real measurable
quantities within linear quantum mechanics~\cite{R1,R2,R221}.
However, one of the major milestones (and a principal thrust
of recent activity) regarding the physical/experimental
realizability of the corresponding Hamiltonians stemmed from progress
in optics both at the theoretical~\cite{ziad,Muga} and at the
experimental~\cite{salamo,dncnat} level. In particular, the realization
that in optics, the ubiquitous loss can be counter-acted by an overwhelming
gain, in order to create a $\mathcal{PT}$-symmetric setup e.g. in a
waveguide dimer~\cite{dncnat} paved the way for numerous developments
especially so at the level of nonlinear systems, as several
researchers studied
nonlinear stationary states, stability and dynamics
of few site configurations \cite{Dmitriev,Li,Guenter,Ramezani,suchkov,Sukhorukov,ZK}
as well as of infinite lattices \cite{Pelin1,Sukh,zheng}.

Interestingly, most of this nonlinear activity has been centered
around Schr{\"o}dinger type systems and for good reason, since
the original proposal by Bender involved quantum mechanical
settings, where this is natural and in addition the optics
proposal was placed chiefly on a similar footing (i.e., the
Schr{\"o}dinger model as paraxial approximation to the Maxwell
equations). Nevertheless, there have been a few notable
exceptions where nonlinear oscillator models (involving
second order differential equations in time) have been
considered. Perhaps the most relevant example, also for the
considerations presented herein, has involved
the realization of  $\cP\cT$-symmetric dimers in the
context of electrical circuits; for the first
work in this context, see~\cite{tsampas}, while that and follow-up activity has
recently been summarized in a review~\cite{tsamp_rev}.
Chiefly, the experimental considerations of these works
focused on the linear variant of the gain-loss oscillator
system. More recently, nonlinear variants of $\cP\cT$-symmetric dimers
in the form of a chain have been proposed in the context
of magnetic metamaterials and in particular for systems
consisting of split-ring resonators~\cite{tsironis}.
The latter setting, while nonlinear is also far more
complex (involving external drive and nonlinear couplings
between adjacent sites) and hence was tackled for the nonlinear
model chiefly at the level of direct numerical simulations.

Our aim herein is to provide a simple, {\it prototypical}
nonlinear model, whose linear analog is effectively the one
used in the experimental investigations of~\cite{tsampas}.
Yet, the nonlinear structure is such that it allows to obtain
a detailed numerical and even considerable analytical insights
on the phenomenology of such a nonlinear $\cP\cT$-symmetric oscillator
dimer. In particular, after formulating and briefly analyzing the linear
$\cP\cT$-symmetric coupled oscillator model,
 we incorporate into it a local cubic nonlinearity (which can, in general, be
of soft or hard form i.e., bearing a prefactor of either sign).
This type of potential, especially in
its bistable form, is well
known to be a canonical example of relevance to numerous
physical settings, including (but not limited to)
phase transitions, superconductivity, field theories and high
energy/particle physics;
see e.g.~\cite{campbell} and the review~\cite{belova}, as well
as references therein. For the resulting $\cP\cT$-symmetric, coupled
nonlinear oscillator model, we provide a detailed analysis of the
existence and stability of breathing (i.e., time-periodic) states
in the system. We focus particularly on symmetric and anti-symmetric
such states that arise from the linear limit of
the problem. We observe that symmetry-breaking type
bifurcations can arise for both the symmetric and anti-symmetric
branches and eventually highlight a nonlinear analog of
$\cP\cT$ phase transition whereby the two branches terminate
hand-in-hand in a saddle-center bifurcation.

To provide an analytical insight into the above results, we use
the Rotating Wave Approximation (RWA) which approximates the
system by the corresponding nonlinear Schr{\"o}dinger type
$\cP\cT$-symmetric dimer for which everything can be solved analytically,
including the stationary states, the symmetry breaking bifurcations
and even the full dynamics~\cite{Ramezani,Li,Pelinov,hadiadd}. A direct
comparison of the RWA-derived Schr{\"o}dinger dimer reveals
natural similarities, but also significant differences between
the two models. For instance, in the way of similarities,
both models bear symmetric
and anti-symmetric branches of solutions, both models bear
principal symmetry-breaking (of the symmetric branch in the
soft --attractive-- nonlinearity and of the anti-symmetric one for the hard
--repulsive-- nonlinearity) bifurcations, and both have $\cP\cT$ phase
transitions, involving the collision and disappearance of
these two branches. On the other hand, in terms of
substantial differences, it is naturally expected that
for soft nonlinearities, the oscillator model can escape
the potential well (and hence have collapse features) even when
the Schr{\"o}dinger model cannot; see for a detailed recent
discussion of such features in the Hamiltonian limit the
work of~\cite{escape}. More importantly for our purposes,
another significant difference is that it turns out that
{\it both} branches, namely both the symmetric and the
anti-symmetric one have destabilizing symmetry-breaking bifurcations,
even though in the Schr{\"o}dinger reduction, only one
of the two branches (the symmetric for soft and the anti-symmetric
for hard, as mentioned above) sustains such bifurcations.
It should also be mentioned that after completing the examination
of the prototypical time-periodic states and their Floquet theory
based stability, we corroborate our bifurcation results by means
of direct numerical simulations in order to explore the different
dynamical evolution possibilities that arise in this system.
These include among others the indefinite growth for the hard
potential and the finite time blow up for the soft potential.

Our presentation of the two models and their similarities/differences
proceeds as follows. In section II, we briefly discuss the
underlying linear model and the prototypical nonlinear extension thereof.
In section III, we discuss the numerical setup for analyzing the
model and its solutions, while in section IV, we provide a
means of theoretical analysis in the form of the rotating
wave approximation. In section V, we present the numerical
results, separating the cases of the soft and hard potential.
Finally, in section VI, we summarize our findings and
present our conclusions, as well as some directions for future study.

\section{Model equations and linear analysis}

We consider the system motivated by recent experimental realizations
in electrical circuits of the form:
\begin{eqnarray}
\ddot{u} &=&-\omega_0^2 u + s v + \gamma \dot{u} ,
 \label{eqn1}
\\
\ddot{v} &=&-\omega_0^2 v + s u - \gamma \dot{v} .
\label{eqn2}
\end{eqnarray}
Here, $\omega_0$ characterizes the internal oscillator at
each mode; in the case of the electrical circuit model
this is the oscillation of each of the charges within the
dimer~\cite{tsampas}. The term proportional to $s$ reflects
the coupling between the two elements in the dimer,
while $\gamma$ is proportional to the amplification/resistance
within the system.

One can try to identify the eigenvalues of the system by using
$u=A e^{i \Omega t}$ and $v=B e^{i \Omega t}$, but one obtains in this
case a quadratic pencil for the relevant eigenvalue problem. It is thus
easier to formulate this as a $4 \times 4$ first order (linear) dynamical
system according to:
\begin{eqnarray}
\dot{u} &=& p ,
\label{eqn3}
\\
\dot{p} &=& - u + \gamma p + s v ,
\label{eqn4}
\\
\dot{v} &=& q ,
\label{eqn5}
\\
\dot{q} &=& s u - v - \gamma q ,
\label{eqn6}
\end{eqnarray}
(where $\omega_0$ has been rescaled without loss of generality to unity,
and other quantities such as $s$ and $\gamma$ and time have
been rescaled by $\omega_0^2$ -the first- and $\omega_0$ -the latter
two-, respectively).
We can then seek solutions of the form $u=A e^{\lambda t}$,
$p=B e^{\lambda t}$, $v=C e^{\lambda t}$ and $q=D e^{\lambda t}$, to
obtain a first order eigenvalue problem which yields the following
eigenvalues
\begin{eqnarray}
\lambda = \pm \frac{\sqrt{-2 + \gamma^2 \pm \sqrt{4 s^2 -4 \gamma^2 + \gamma^4}}}{\sqrt{2}} .
\label{eqn7}
\end{eqnarray}
These two pairs of imaginary (for small $\gamma$) eigenvalues will
collide and give rise to a quartet for $\gamma > \gamma_{PT}$, where
$\gamma_{PT}$ satisfies the condition:
\begin{eqnarray}
\gamma^4 - 4 \gamma^2 + 4 s^2=0.
\label{eqn8}
\end{eqnarray}
Hence Eq.~(\ref{eqn8}) will define, {\it at the linear level},
the point of the so-called~\cite{R1,R2,R221} $\cP\cT$ phase transition
and of bifurcation into the complex plane.

Now, our main interest in what follows will be
to examine a prototypical nonlinear variant  of the problem
which will be formulated as follows.
In particular, we set up the form of the equations:
\begin{eqnarray}
\ddot{u} &=&- u + s v + \gamma \dot{u} + \epsilon u^3 ,
 \label{eqn9}
\\
\ddot{v} &=&- v + s u - \gamma \dot{v} + \epsilon v^3 .
\label{eqn10}
\end{eqnarray}
Here, in parallel to what is done in the $\cP\cT$-symmetric
Schr{\"o}dinger dimer typically~\cite{dncnat,Ramezani,Li}, we
have added a cubic onsite nonlinearity on each one of the nodes.
For $\epsilon>0$, this nonlinearity is soft, imposing a finite
(maximal energy) height type of potential, enabling the possibility
of indefinite growth by means of the escape scenario considered
earlier e.g. in~\cite{escape} (see also references therein).
On the other hand, for $\epsilon <0$,
the nonlinearity is hard, and the potential is monostable
bearing only the ground state at $0$ and no possibility for
such finite time collapse (in the Hamiltonian analog of the model,
only the potential for oscillations around the $0$ state exists
in this case of the hard potential).

We now discuss the setup and numerical methods, as well as the
type of diagnostics that we use for this system. In our
description below, we follow an approach reminiscent of
that in~\cite{phason}.

\section{Setup, Diagnostics and Numerical Methods}

\subsection{Existence of Periodic Orbit Solutions}

In order to calculate periodic orbits in the $\cP\cT$ nonlinear oscillator
dimer, we make use of a Fourier space implementation of the dynamical
equations and continuations in frequency or gain/loss parameter are performed
via a path-following (Newton-Raphson) method. Fourier space methods are based
on the fact that the solutions are $\Tb$-periodic; for a detailed explanation of these methods, the reader is referred to Refs.~\cite{AMM99,Marin,Cuevas}.
The method has the advantage, among others, of providing an explicit,
analytical form of the Jacobian. Thus, the solution for the two nodes
can be expressed in terms of a
truncated Fourier series expansion:

\begin{equation}\label{eq:series}
    u(t)=\sum_{k=-k_m}^{k_m} y_k\exp(\ii k \wb t),\qquad v(t)=\sum_{k=-k_m}^{k_m} z_k\exp(\ii k \wb t) ,
\end{equation}

\noindent with $k_m$ being the maximum of the absolute value of the running
index $k$ in our Galerkin truncation of the full Fourier series
solution. In the numerics, $k_m$ has been chosen as 21. After the
introduction of (\ref{eq:series}), the dynamical equations
(\ref{eqn9})-(\ref{eqn10}) yield a set of $2\times(2k_m+1)$
nonlinear, coupled algebraic equations:

\begin{eqnarray}\label{eq:Fourier}
    F_{k,1} &\equiv& -\wb^2k^2y_k-\ii\gamma\wb k y_k+\F_k[V'(u)]-s z_k=0 , \\
    F_{k,2} &\equiv& -\wb^2k^2z_k+\ii\gamma\wb k z_k+\F_k[V'(v)]-s y_k=0 ,
\end{eqnarray}

\noindent with $V'(u)=u-\epsilon u^3$. Here, $\F_k$ denotes the Discrete Fourier Transform:

\begin{equation}
    \F_k[V'(u)]=\frac{1}{N}\sum_{n=-k_m}^{k_m}V'\left(\sum_{p=-k_m}^{k_m}y_p\exp\left[\ii \frac{2\pi p n}{N}\right]\right)
    \exp\left[-\ii \frac{2\pi k n}{N}\right],
\end{equation}
with $N=2k_m+1$. The procedure for $\F_k(v)$ is similar to the previous one.
As $u(t)$ and $v(t)$ must be real functions, it implies that $y_{-k}=y^*_k,\ z_{-k}=z^*_k$.

An important diagnostic quantity for probing the dependence of
the solutions on parameters such as the gain/loss strength
$\gamma$, or the oscillation frequency $\wb$
is the averaged over a period energy, defined as:
\begin{equation}
    <H>=\frac{1}{T}\int_0^T\ H(t)\,\mathrm{d}t ,
\end{equation}
with the Hamiltonian (of the case without gain/loss) being
\begin{equation}
    H=\frac{\dot{u}^2+\dot{v}^2+u^2+v^2}{2}-\frac{\epsilon}{4}(u^4+v^4)-suv,
\end{equation}
and constituting a conserved quantity of the dynamics in the Hamiltonian
limit of $\gamma=0$.

\subsection{Linear stability equations}

In order to study the spectral stability of periodic orbits, we introduce a small perturbation $\{\xi_1,\xi_2\}$ to a given solution
$\{u_0,v_0\}$ of Eqs. (\ref{eqn9})-(\ref{eqn10}) according to $u=u_{0}+\xi_1$, $v=v_0+\xi_2$. Then, the equations satisfied to first order in $\xi_n$ read:

\begin{eqnarray}\label{eq:stab}
    \ddot\xi_1+V''(u_0)\xi_1-\gamma \dot\xi_1-s\xi_2 &=& 0 ,\\
    \ddot\xi_2+V''(v_0)\xi_2+\gamma \dot\xi_2-s\xi_1 &=& 0 ,
\end{eqnarray}

\noindent or, in a more compact form: $\mathcal{N}(\{u(t),v(t)\})\xi=0\,$, where $\mathcal{N}(\{u(t),v(t)\})$ is the relevant linearization operator.
In order to study the spectral (linear) stability analysis of the relevant
solution, a Floquet analysis can be performed if there exists $T_b\in\mathbb{R}$ so that the map $\{u(0),v(0)\}\rightarrow \{u(\Tb),v(\Tb)\}$ has a fixed
point (which constitutes a periodic orbit of the original system).
Then, the stability
properties are given by the spectrum of the Floquet operator $\mathcal{M}$ (whose matrix representation is the monodromy) defined as:
\begin{equation}\label{eq:monodromy}
    \left(\begin{array}{c} \{\xi_{n}(\Tb)\} \\ \{\dot\xi_{n}(\Tb)\} \\ \end{array}
    \right)=\mathcal{M}\left(\begin{array}{c} \{\xi_{n}(0)\} \\ \{\dot\xi_{n}(0)\} \\ \end{array}
    \right) .
\end{equation}

The $4\times4$ monodromy eigenvalues $\Lambda=\exp(\ii\theta)$ are dubbed the {\em Floquet multipliers} and $\theta$ are denoted as {\em Floquet exponents}
(FEs). This operator is real, which implies that there is always a pair of
multipliers at $1$ (corresponding to the so-called phase and growth modes~\cite{Marin,Cuevas})
and that the eigenvalues come in pairs $\{\Lambda,\Lambda^*\}$.
As a consequence, due to the ``simplicity'' of the FE structure (one
pair always at $1$ and one additional pair) there cannot exist Hopf
bifurcations in the dimer, as such bifurcations would
imply the collision of two pairs of multipliers and the consequent formation
of a quadruplet of eigenvalues which is impossible here. Nevertheless,
in the present problem, the motion of the pair of multipliers
can lead to instability through exiting (through $1$ or $-1$) on the
real line leading to one multiplier (in absolute value) larger
than $1$ and one smaller than $1$. We will explore a scenario of
this kind of instability in what follows.

Having set up the existence and stability problem, we now
complete our theoretical analysis by exploring the outcome
of the Rotating Wave Approximation (RWA).

\section{An Analytical Approach: The Rotating Wave Approximation}

The RWA provides a means of connection with the extensively analyzed
$\cP\cT$-symmetric Schr{\"o}dinger dimer~\cite{Ramezani,Li,dwell}.
This link follows a path similar to what has been earlier proposed
e.g. in \cite{RWA1,RWA2,Morgante}. In particular,
the following ansatz is used to approximate the solution
of the periodic orbit problem as a roughly monochromatic wavepacket
of frequency $\wb$ (for $\phi_{1,2}$ in what follows we will seek
stationary states).
\begin{equation}\label{eq:RWA0}
    u(t)\approx \phi_1(t)\exp(\ii \wb t)+\phi^*_1(t)\exp(-\ii \wb t),\qquad v(t)\approx \phi_2(t)\exp(\ii \wb t)+\phi^*_2(t)\exp(-\ii \wb t) .
\end{equation}

By supposing that $\dot\phi_n\ll\wb\phi_n$ and $\ddot\phi_n\ll\wb\dot\phi_n$
(i.e., that $\phi$ varies slowly on the scale of the oscillation of
the actual exact time periodic state),
discarding the terms multiplying $\exp(\pm 3\ii\wb t)$, the dynamical equations (\ref{eqn9})-(\ref{eqn10}) transform into a set of coupled
Schr{\"o}dinger type equations:
\begin{eqnarray}\label{eq:DNLS}
2\ii\wb\dot{\phi}_1 &=& [(\wb^2-1)+3\epsilon|\phi_1|^2+\ii\wb\gamma]\phi_1 + s \phi_2 ,
\nonumber
\\
2\ii\wb\dot{\phi}_2 &=& [(\wb^2-1)+3\epsilon|\phi_2|^2-\ii\wb\gamma]\phi_2 + s \phi_1,
\end{eqnarray}
i.e., forming, under these approximations, a
$\cP\cT$-symmetric Schr{\"o}dinger dimer. The stationary solutions of
this dimer can then be used in order to reconstruct via
Eq.~(\ref{eq:series}) the solutions of the RWA to the original
$\cP\cT$-symmetric oscillator dimer. These stationary solutions
for $\phi_1(t)\equiv y_1$ and $\phi_2(t)\equiv z_1$ satisfy the
algebraic conditions

\begin{eqnarray}\label{eq:RWA1}
    Ey_1&=&\kappa z_1+|y_1|^2y_1+\ii\Gamma y_1 , \\
\label{eq:RWA2}
    Ez_1&=&\kappa y_1+|z_1|^2z_1-\ii\Gamma z_1 ,
\end{eqnarray}

with

\begin{equation}
    E=\frac{1-\wb^2}{3\epsilon},\quad \kappa=\frac{s}{3\epsilon},\quad \Gamma=\frac{\gamma\wb}{3\epsilon}
\end{equation}

Recast in this form, Eqs.~(\ref{eq:RWA1})-(\ref{eq:RWA2}) are
identical to that in \cite[Eq.(6)]{dwell}. We express $y_1$ and $z_1$ in polar form:

\begin{equation}
    y_1=A\exp(\ii \theta_1),\qquad z_1=B\exp(\ii \theta_2),\qquad \varphi=\theta_2-\theta_1 ,
\end{equation}
and then rewrite the stationary equations as

\begin{eqnarray}\label{eq:RWAb1}
    EA&=&\kappa B \cos(\varphi) + A^3 ,
\\
\label{eq:RWAb2}
    EB&=&\kappa A \cos(\varphi) + B^3 ,
\\
\label{eq:RWAb3}
\sin(\varphi) &=& - \Gamma A/(\kappa B) = - \Gamma B/(\kappa A) .
\end{eqnarray}

In the Hamiltonian case $\gamma=0$, and consequently, $\sin\varphi=0$. Three different solutions may exist therein, namely, the symmetric (S), anti-symmetric (A) and asymmetric (AS) solutions, given by:

\begin{eqnarray}\label{eq:RWAgamma0}
    A=B , & A^2=E-\kappa=\frac{1-\wb^2-s}{3\epsilon} , & \textrm{S solution} \\
    A=-B , & A^2=E+\kappa=\frac{1-\wb^2+s}{3\epsilon} , & \textrm{A solution} \\
    B=\kappa/A=s/(3\epsilon A) , & A^2=(E\pm\sqrt{E^2-4\kappa^2})/2=\left[(1-\wb^2)\pm\sqrt{(1-\wb^2)^2-4s^2}\right]/(6 \epsilon) . & \textrm{AS solution}
\end{eqnarray}

The symmetric solution derives from the linear mode located at $\omega_S=\sqrt{1-s}$ whereas the anti-symmetric solution bifurcates from the mode at $\omega_A=\sqrt{1+s}$. Straightforwardly (by examining the quantity under the
radical in its profile), the AS solution exists for $\wb\leq\sqrt{1-2s}$,
bifurcating via a symmetry-breaking pitchfork bifurcation
from the S solution if the potential is soft ($\epsilon>0$).
On the contrary, if the potential is hard, the AS solution
bifurcates from the A solution and exists for $\wb\geq\sqrt{1+2s}$.
The emerging (``daughter'') AS solutions inherit the stability of
their S or A ``parent'' and are therefore stable whereas the respective
parent branches become destabilized past the bifurcation point.

For $\gamma\neq0$, the AS is no longer a stationary solution and only S and A
solutions exist as exact stationary states in the $\cP\cT$-symmetric
Schr{\"o}dinger dimer [as is directly evident e.g. from
Eq.~(\ref{eq:RWAb3})]. These solutions have $A=B$ taking the values:
\begin{eqnarray}
    A &= \left[\left(1-\wb^2-\sqrt{s^2-\gamma^2\wb^2}\right)/(3\epsilon)\right]^{1/2} , & ~~~\textrm{S solution} \\
    A &= \left[\left(1-\wb^2+\sqrt{s^2-\gamma^2\wb^2}\right)/(3\epsilon)\right]^{1/2} & ~~~ \textrm{A solution}
\end{eqnarray}
and $\sin\varphi=-\Gamma/\kappa$. As $|\sin\varphi|\leq1$, the
solutions must fulfill $\gamma\leq\gamma_{SC}$, with $\gamma_{SC}=\frac{s}{\wb}$ i.e., there is a saddle-center bifurcation (namely, the RWA
predicted nonlinear analog
of the $\cP\cT$ phase transition) taking place $\gamma=\gamma_{SC}$.

The average energy for both the S and A solutions is given by the same expression:

\begin{equation}\label{eq:RWAenergy}
    <H>\approx 4\wb^2A^2+3\epsilon A^4 .
\end{equation}

As both solutions coincide at the $\cP\cT$ bifurcation critical point, their energy will also be the same therein.

We now turn to the linear stability of the different solutions
within the RWA. The spectral analysis of the S and A solutions
can be obtained by considering small perturbations [of order ${\rm O}(\delta)$, with $0< \delta \ll 1$] of the stationary solutions.
The stability can be determined by substituting the ansatz below into (\ref{eq:DNLS}) and then solving the ensuing [to O$(\delta)$] eigenvalue problem:

\begin{eqnarray}
\phi_1(t) &=& y_1 + \delta (a_1 e^{-\ii\theta t/T} + b^{*}_1 e^{\ii\theta^* t/T}) , \nonumber \\
\phi_2(t) &=& z_1 + \delta (a_2 e^{-\ii\theta t/T} + b^{*}_2 e^{\ii\theta^* t/T}) ,
\end{eqnarray}
with $T=2\pi/\wb$ being the orbit's period and $\theta$ being the Floquet exponent (FE). The non-zero FEs are given by:

\begin{eqnarray}
    \theta &=\pm\frac{2\pi}{\wb^2}\left[2\left(s^2-\gamma^2\wb^2\right)-\left(1-\wb^2\right)\sqrt{s^2-\gamma^2\wb^2}\right]^{1/2}  , & ~~~ \textrm{S solution} \\
    \theta &=\pm\frac{2\pi}{\wb^2}\left[2\left(s^2-\gamma^2\wb^2\right)+\left(1-\wb^2\right)\sqrt{s^2-\gamma^2\wb^2}\right]^{1/2} . & ~~~ \textrm{A solution}
\end{eqnarray}

As instability is marked by an imaginary value of $\theta$, the above expression implies that there is a stability change when the square root argument becomes zero, i.e. $\gamma=\gamma_\textrm{stab}$, with

\begin{equation}
    \gamma_\textrm{stab}=\frac{\sqrt{4s^2-(1-\wb^2)^2}}{2\wb}.
\end{equation}

A straightforward analysis shows, in addition, that the S (A) solution experiences the change of stability bifurcation when $\wb$ is smaller (larger) than 1. Thus, the S (A) solution is always stable when the potential is hard (soft) and stable if $\gamma<\gamma_\textrm{stab}$ when the potential is soft (hard).
As we will see below, it is precisely this prediction of the RWA
that will be ``violated'' from the full numerics of the $\cP\cT$-symmetric
oscillator model. In particular, it will be found that in the latter model,
{\it both} branches in each (soft or hard)
case can become unstable through such
symmetry-breaking bifurcations within suitable parametric regimes.

As a final theoretical remark, it is relevant to point out that
while no additional stationary solutions have been argued to
exist in the Schr{\"o}dinger dimer, a ``reconciliation'' with the
expected picture of a pitchfork bifurcation has been offered
e.g. in~\cite{dwell} (see also references therein) through the
notion of the so-called ghost states. These are solutions for
which the propagation constant parameter $E$ becomes complex
(and the pitchfork bifurcation resurfaces in diagnostics such as
the imaginary part of $E$). Nevertheless and especially because
complex eigenvalue parameters are of lesser apparent physical
relevance in models such as the oscillator one considered
herein, we will not further pursue an analogy to such ghost
states here, but will instead restrain our considerations hereafter
to the S and A branches of time-periodic solutions.

\section{Numerical results and Comparison with the Rotating Wave
Approximation}

In this section, we identify the relevant, previously discussed
periodic orbits by numerically solving in the Fourier space
the dynamical equations set (\ref{eqn9})-(\ref{eqn10}).
We have considered the two cases
of $\epsilon=\pm1$, with $\epsilon=1$ corresponding to the soft potential,
while $\epsilon=-1$ corresponds to the hard potential case.

We analyze the properties of phase-symmetric (S) and phase-anti-symmetric (A) solutions, characterized respectively by the following properties:

\begin{equation}\label{eq:sym}
    u(0)=v(0), \dot u(0)=-\dot v(0)\ \mathrm{(S)} \qquad u(0)=-v(0), \dot u(0)=\dot v(0)\ \mathrm{(A)}
\end{equation}

or, in terms of the Fourier coefficients:

\begin{equation}
    y_k=z_k^*\ \mathrm{(S)} \qquad y_k=-z_k^*\ \mathrm{(A)}
\end{equation}

\begin{figure}[t]
\begin{center}
\begin{tabular}{cc}
    \includegraphics[scale=0.4]{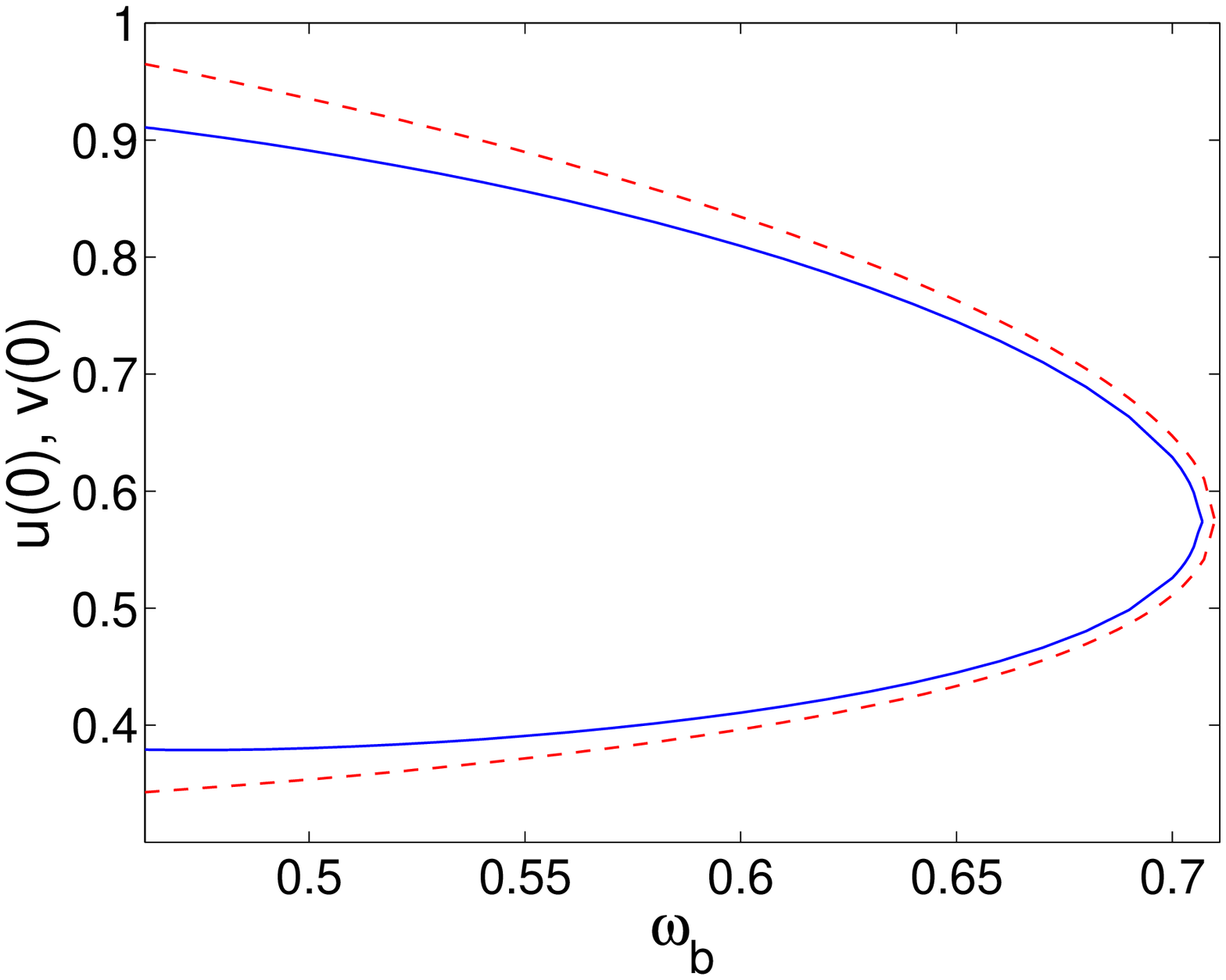} &
    \includegraphics[scale=0.4]{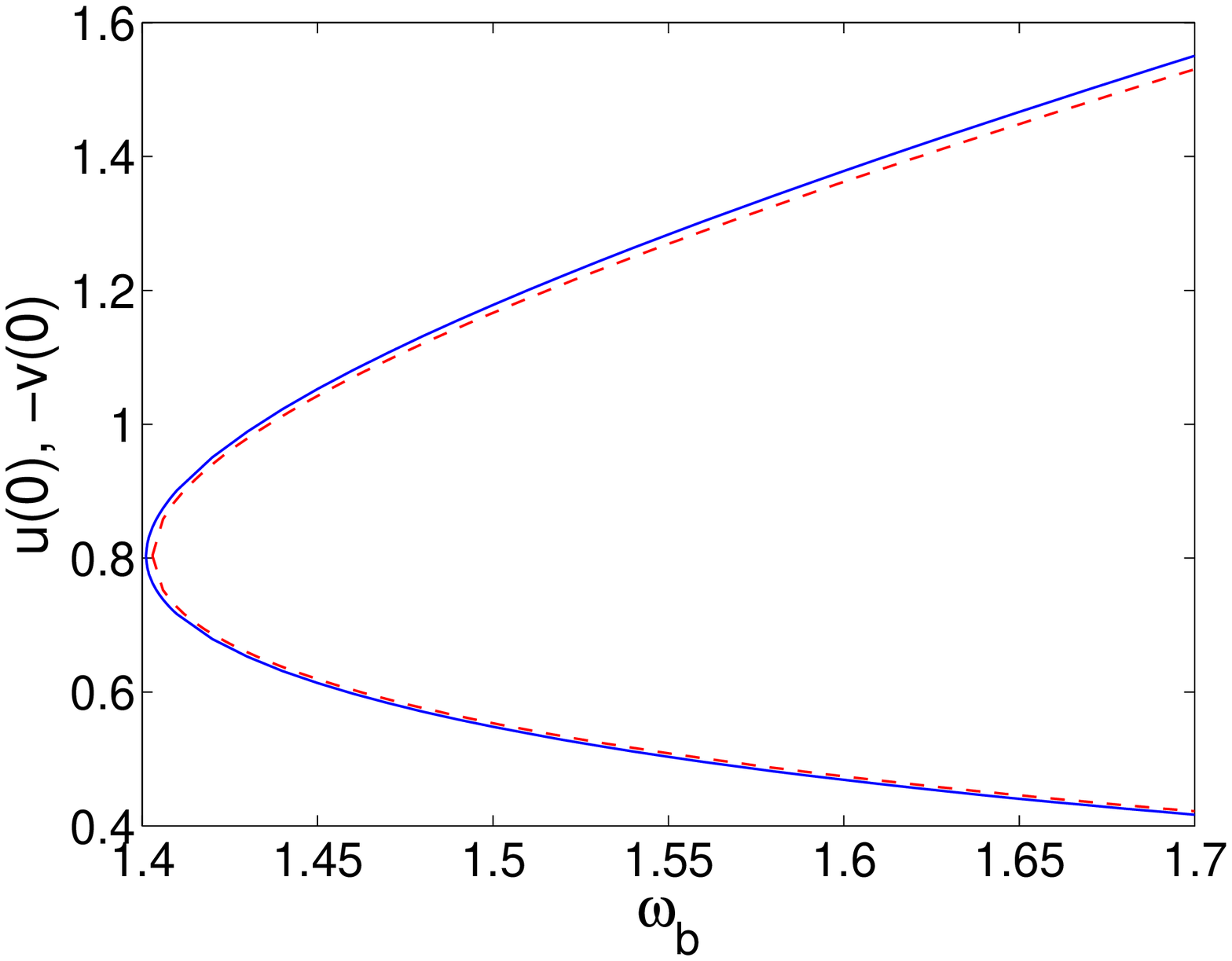} \\
\end{tabular}
\caption{(Color Online)
Amplitude of the couplers for the asymmetric solution in the
Hamiltonian case of $\gamma=0$ and for
$s=\sqrt{63}/32$, $\epsilon=1$ (left panel) and
$s=\sqrt{15}/8$, $\epsilon=-1$ (right panel).
The dashed lines correspond to the predictions of the RWA theory
(see the discussion in the text).
{It is worth noticing that in the left panel, asymmetric solutions are unstable towards finite time blow up for frequencies smaller than those shown in the figure. On the contrary, in the right panel, solutions are
always stable for frequencies higher than those shown in the picture.}}
\label{fig:asym}
\end{center}
\end{figure}

It is worth remarking that for $\gamma=0$, the solutions are
time-reversible and, consequently, we select $\dot u(0)=\dot v(0)=0$.
We recall that in the Hamiltonian limit, the RWA predicts that the S (A) solution becomes unstable at $\wb=\omega_S\equiv\sqrt{1-2 s}$ ($\wb=\omega_A\equiv\sqrt{1+ 2 s}$) in the soft (hard) case, giving rise to an asymmetric (AS) solution.
For the particular case of $s=\sqrt{63}/32$, the results for
the soft case are shown (for $\gamma=0$) in the left
panel of Fig.~\ref{fig:asym}, while those for the hard potential case
are shown in the right panel of the figure (for $s=\sqrt{15}/8$).
The solid lines of the direct numerical computation appear to have
good agreement with the dashed lines of the RWA, as regards the
predicted amplitudes of the asymmetric node equilibrium values, at
least near the bifurcation point. Interestingly, this agreement
is {\it considerably better} in the hard case than in the soft nonlinearity
case. This will be a continuous theme within the results that follow,
i.e., we will see that the hard case is generally very accurately
described by the RWA (even in the presence of gain/loss), while the
soft case is only well approximated sufficiently close to the linear
limit. It should be noted
that a fundamental difference between the nonlinear Schr{\"o}dinger
type dimer and the $\phi^4$ oscillator one is expected as
the amplitude of the solution increases (and so does
the deviation from the symmetry breaking point). In particular,
the former model due to its norm conservation does {\it not}
feature finite time collapse (or any type of infinite
growth for that matter when $\gamma=0$). On the contrary,
the latter model has the potential for finite time collapse when the
amplitude of the nodes exceeds the unit height of the potential
(see a detailed analysis of this ``escape'' phenomenology in
the recent work of~\cite{escape} and references therein).
It is thus rather natural that the two models should
{\it significantly} deviate from each other as this parameter range
is approached.

\subsection{Soft potential}

\begin{figure}[t]
\begin{center}
    \includegraphics[scale=0.4]{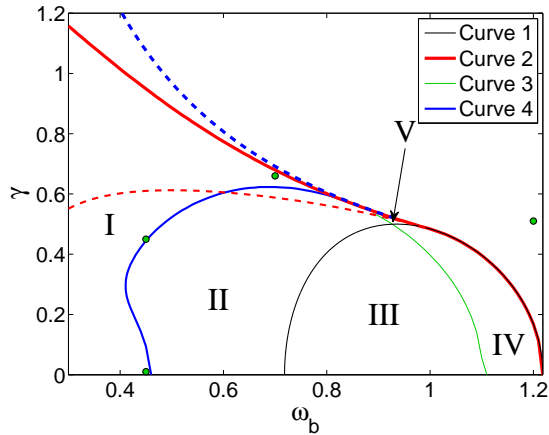}
\caption{(Color Online) Plane with curves and regions of solutions that share the same properties when $\epsilon=1$ and $s=\sqrt{15}/8$ (see text). In the non-labeled region
of the top right, neither S nor A solutions exist. Here Curve 1 corresponds to the linear limit, Curve 2
denotes the $\cP \cT$ phase transition curve, Curve 3 indicates the destabilization of the A branch, while Curve 4
is associated with the destabilization of the symmetric branch. The
detailed description of the different regions enclosed by the
curves is offered in the text. Dashed lines correspond to the RWA predictions. Notice that the colors of the dashed lines are inverted with respect to that of the numerical results for a better visualization.
I.e., the blue dashed line is the theoretical prediction corresponding to the
red solid one, while the red dashed line represents the prediction
corresponding to the blue solid one. This inversion pattern is followed
also in all figures comparing theory and numerical computations from here on.
Dots mark the parameters for which simulations are performed in Fig. \ref{fig:dynamics1}.} \label{fig:softplane}
\end{center}
\end{figure}

We analyze firstly the soft case ($\epsilon=+1$) in the presence now
of the gain/loss term proportional to $\gamma$. Figure~\ref{fig:softplane}
shows a $\gamma$-$\wb$ full two-parameter plane summarizing the existence
properties of the solutions and separating the different regimes thereof.

\begin{itemize}

\item

Curve 1 corresponds to the linear modes, obtained in section II;
The increasing part of the curve fulfills $\omega_1 = \sqrt{1 - \gamma^2/2 - \sqrt{s^2 - \gamma^2 + \gamma^4/4}}$ and corresponds to symmetric linear modes at $\gamma=0$; the decreasing part $\omega_2 = \sqrt{1 - \gamma^2/2 + \sqrt{s^2 - \gamma^2 + \gamma^4/4}}$ holds for the branch stemming from
the anti-symmetric linear modes of $\gamma=0$ [cf. with
Eq.~(\ref{eqn7})]. These two classes of linear modes collide and
disappear hand-in-hand at the value of $\gamma$ predicted by
Eq.~(\ref{eqn8}). For this soft case, solutions are expected
to exist (in analogy with the Schr{\"o}dinger case)
for $\wb < \omega_1$ for the symmetric branch and for
$\wb < \omega_2$ for the anti-symmetric branch
(for a given value of $\gamma$).

\item

Curve 2 corresponds to the $\cP \cT$ phase transition, i.e., at the
nonlinear level it corresponds to the saddle-center bifurcation leading
to the termination of both the A and S branches. Above this curve, there
do not exist any periodic orbits and the system dynamics
generically leads to indefinite growth.
This curve overlaps with Curve 1 for high $\wb$.

\item

{Curves 3 and 4 separate stable and unstable solutions of the anti-symmetric and symmetric branches, respectively.
In particular, they represent the threshold for the symmetry-breaking bifurcation of the corresponding branches}.

\end{itemize}

The regions limited by the curves above are the following ones:

\begin{itemize}

\item Region I: Both solutions S and A are unstable, as they have
both crossed the instability inducing curves 3 and 4.

\item Region II: Solutions S are stable (as they have not crossed
curve 4) whereas A are unstable (as bifurcating from the
decreasing part of curve 1, they have crossed the instability
threshold of curve 3).

\item Region III: Solutions S do not exist (as such solutions only exist
to the left of the increasing part of curve 1)
and solutions A are unstable (as they have crossed the instability
threshold of curve 3).

\item Region IV: Solutions S do not exist (for the same
reason as in III) and solutions A are stable. I.e., they are stable
between their bifurcation point --decreasing part of curve 1-- and
instability threshold --curve 3--.

\item Region V: {Stable A and S solutions coexist in this narrow
region prior to their termination in the saddle-center bifurcation
occurring on Curve 2.}

\end{itemize}

\begin{figure}[t]
\begin{center}
\begin{tabular}{cc}
    \includegraphics[scale=0.4]{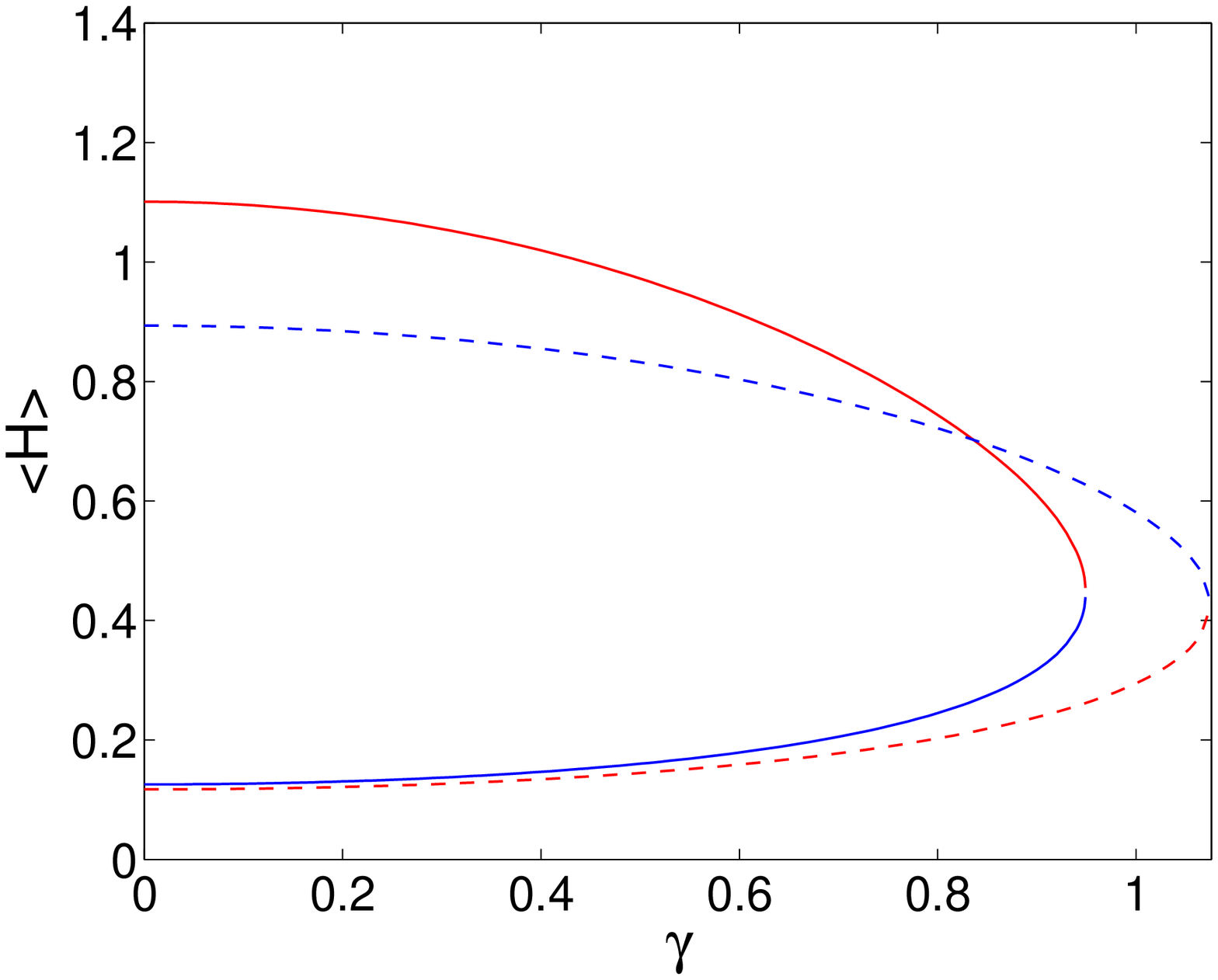} &
    \includegraphics[scale=0.4]{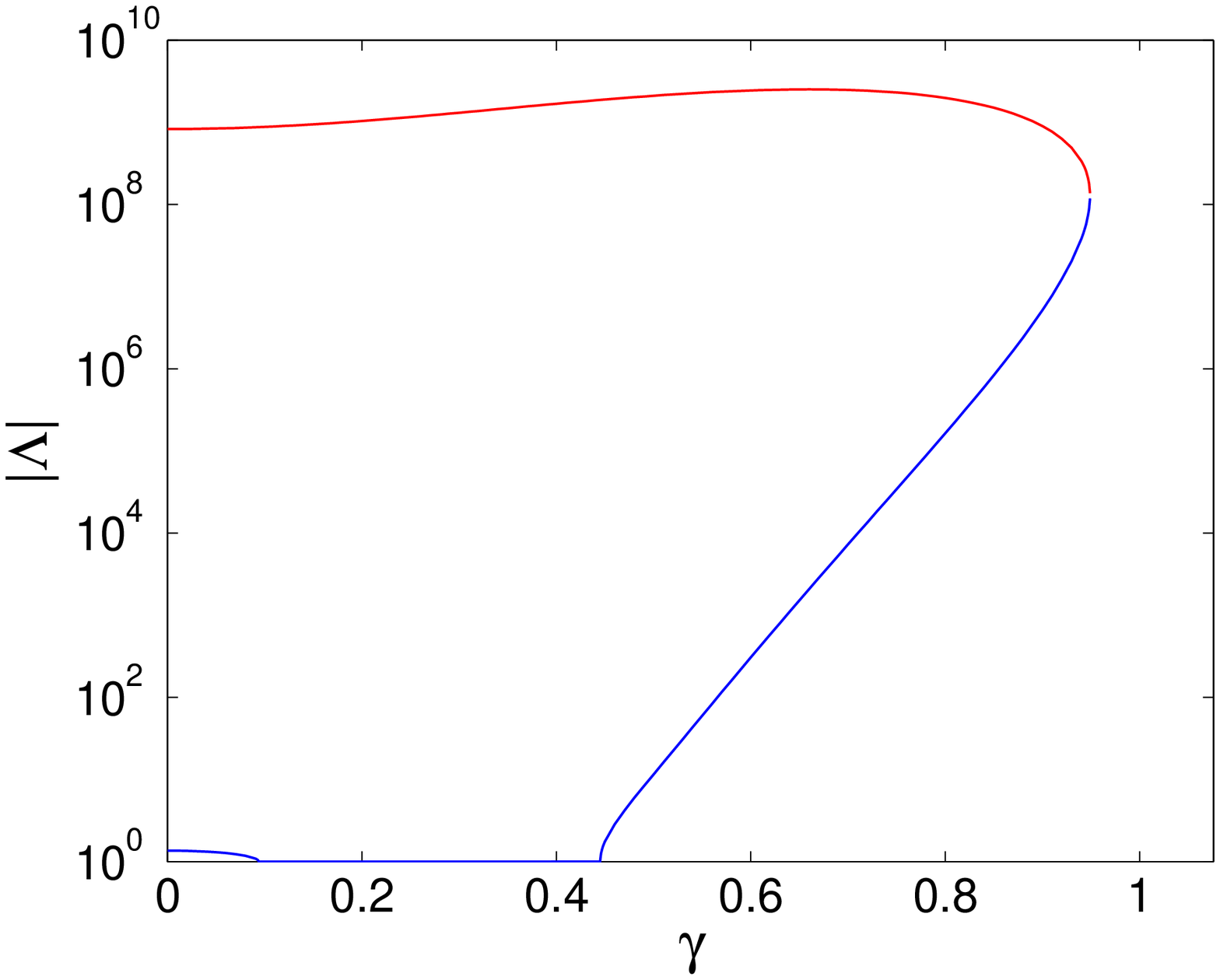} \\
    \includegraphics[scale=0.4]{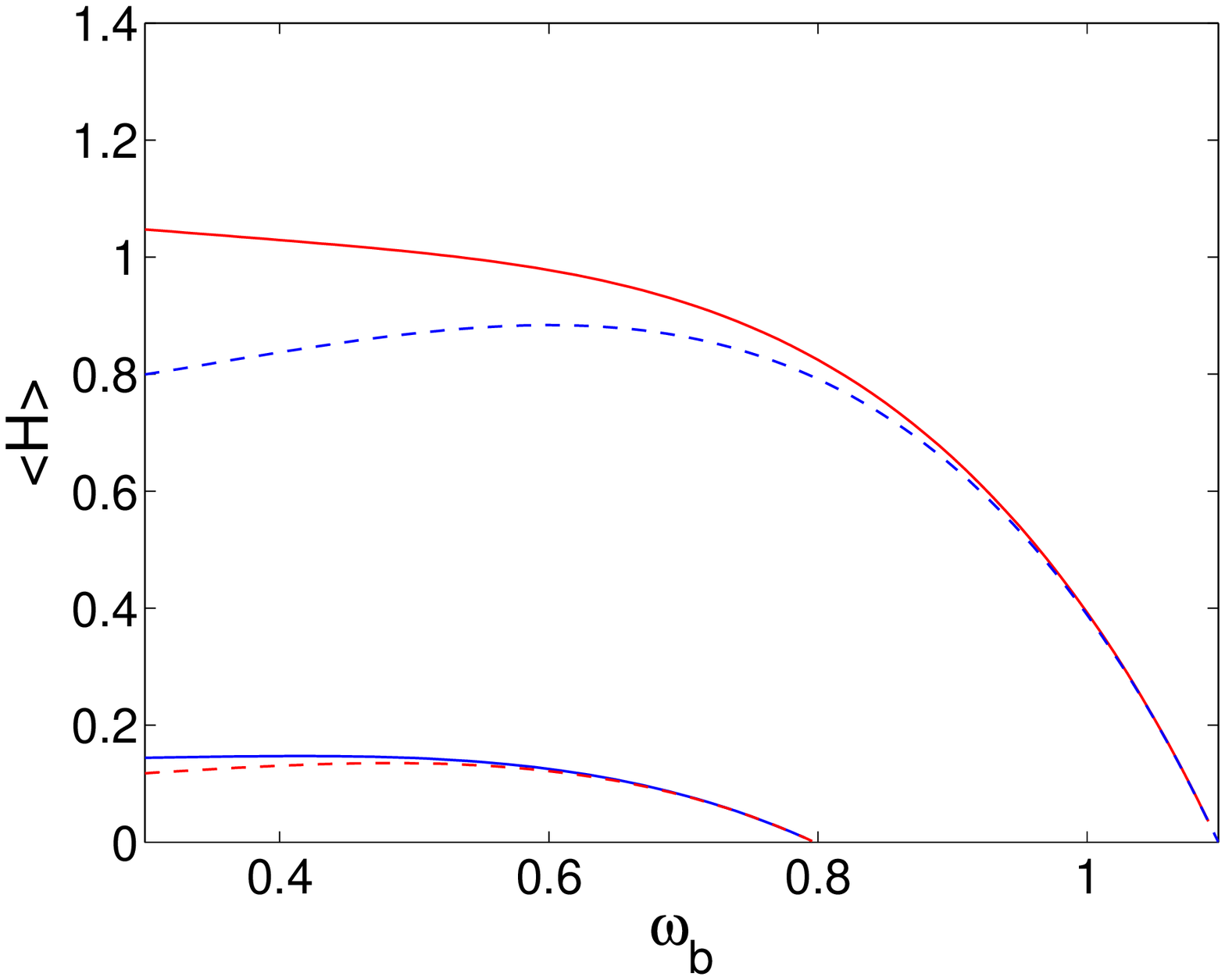} &
    \includegraphics[scale=0.4]{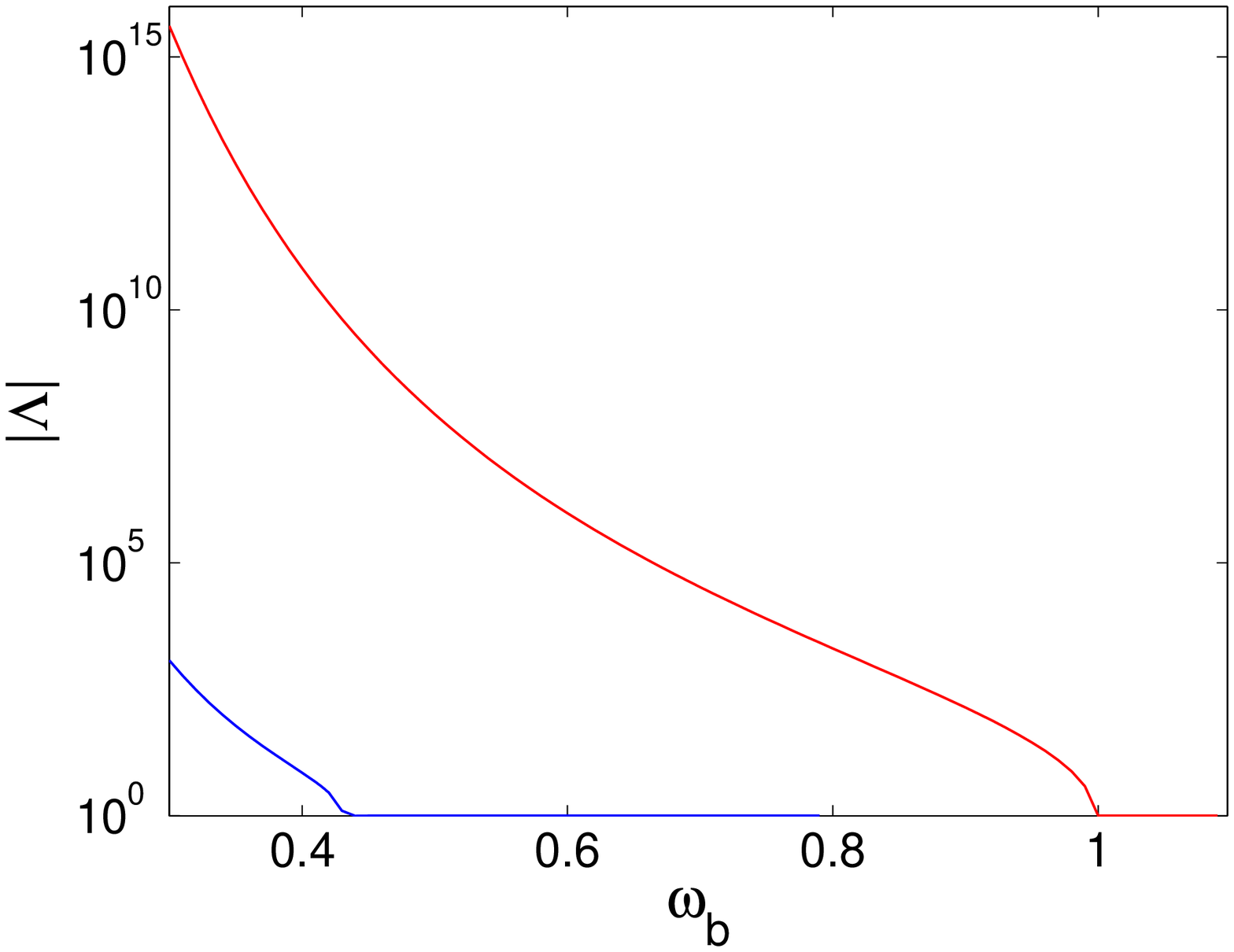} \\
\end{tabular}
\caption{(Color Online)
Averaged energy (top left) and modulus of the Floquet multipliers
(top right)
[only the multipliers with moduli higher than one are shown] as a function
of the gain/loss parameter $\gamma$ for $\epsilon=1$, $\wb=0.45$,
and $s=\sqrt{15}/8$. Notice the logarithmic scale in the
$y$-axis of the latter graph. In the left panel,
 blue (red) solid line corresponds to the S (A) solution, while
red (blue), i.e., reversed colors, dashed lines correspond to the S (A)
branch RWA predictions. One can observe a reasonable agreement between
the numerical energy and its theoretical prediction counterpart,
although some discrepancy is clearly visible for this low (i.e., far
from the linear limit) value of the frequency. The bottom panel
shows similar comparisons but now for the dependence on
$\wb$ for fixed $\gamma=0.4$. Here it is evident that the agreement is good
close to the linear limit of larger values of $\wb$ and progressively
worsens as one departs from this limit (by lowering the breather frequency).}
\label{fig:examp1}
\end{center}
\end{figure}

In Fig.~\ref{fig:examp1}, some typical examples of mono-parametric
continuations of the relevant solutions are given. The top panels
illustrate continuations as a function of the gain/loss
parameter $\gamma$ for a fixed value of the frequency
$\wb=0.45$, while the bottom ones illustrate a continuation
as a function of $\wb$ for a given value of $\gamma=0.4$.
The comparison of the numerically obtained symmetric and
anti-symmetric solutions with the ones obtained analytically
by virtue of the RWA (in reverse colors-- see the figure) is also
offered. It can be inferred that generally the RWA offers
reasonable qualitative match to the numerically exact, up to
a prescribed tolerance, solutions, although clearly quantitative
comparison is less good, at least for the low (i.e., far
from the linear limit) value of $\wb$ in the top panels.
This deficiency of the method (explained also previously)
as one departs far from the linear limit is more clearly
illustrated in the $\wb$ dependence. Close to the limit,
the RWA does an excellent job of capturing both branches,
but things become progressively worse as $\wb$ decreases.
Furthermore, as discussed above the right panels showcase
a stability change as occurring for {\it both} branches, while
the RWA predicts a destabilization solely of the symmetric
branch for this soft nonlinearity case.

\begin{figure}[t]
\begin{center}
\begin{tabular}{cc}
\includegraphics[scale=0.4]{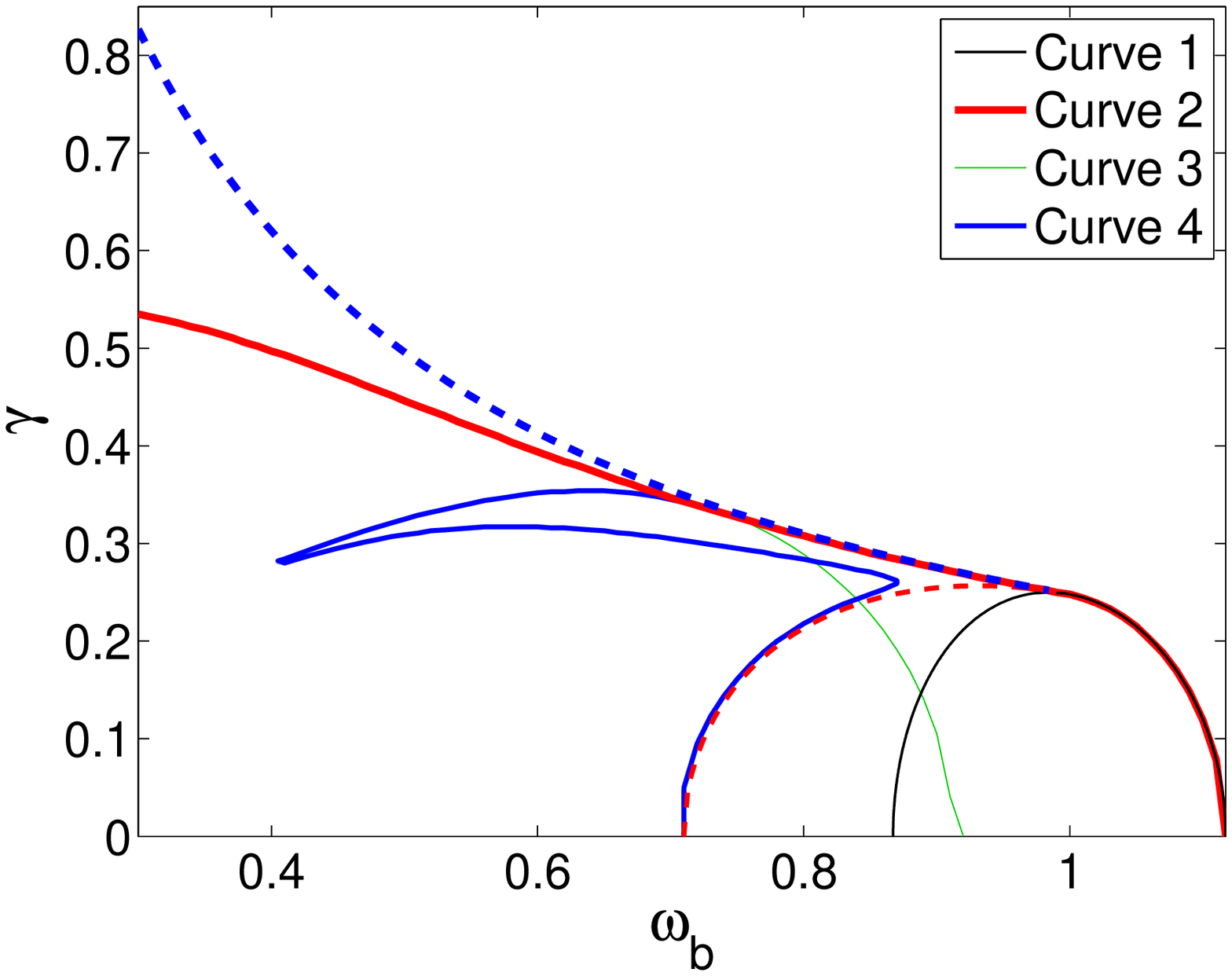} \\
    \includegraphics[scale=0.4]{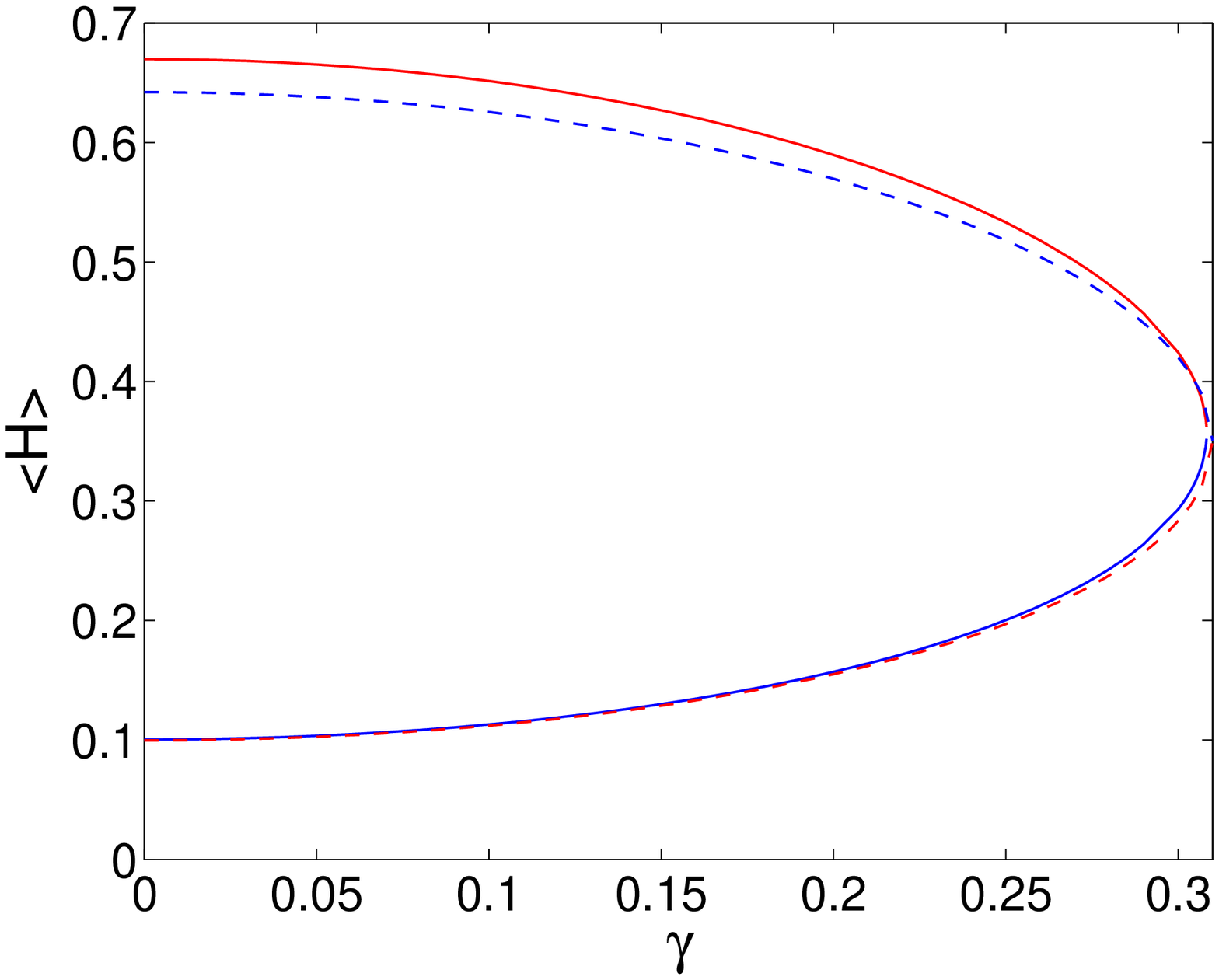} &
    \includegraphics[scale=0.4]{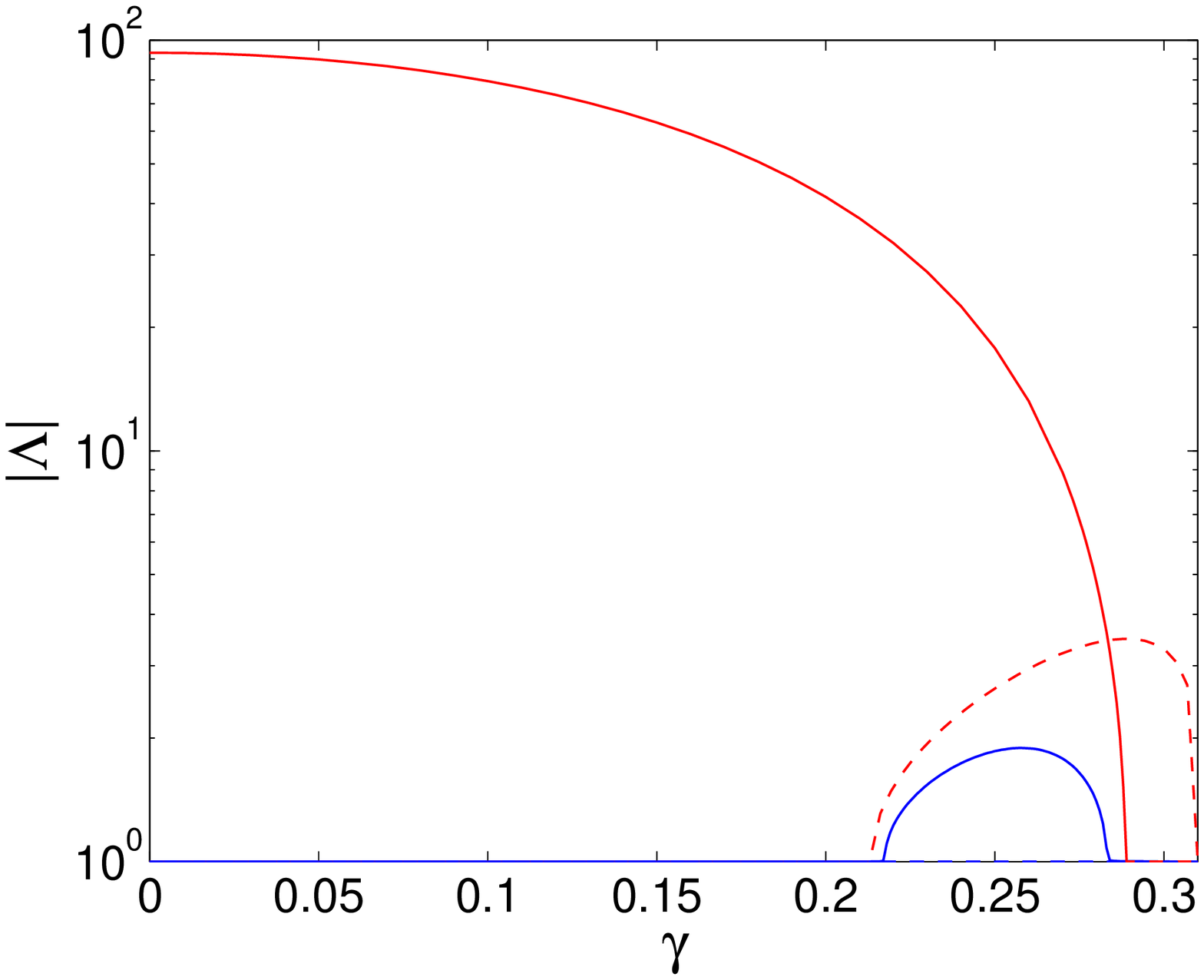} \\
\end{tabular}
    \caption{(Color Online) Same as the Fig.~\ref{fig:softplane} (top panel) and as the
top panel of Fig.~\ref{fig:examp1} (bottom panels) but for the soft
nonlinearity case of lower coupling $s=\sqrt{63}/32$. Again the
full numerical results are offered by the solid lines for the
two branches (blue for S and red for A), while the dashed lines
with reverse colors correspond to the RWA.
} \label{fig:softplane2}
\end{center}
\end{figure}

In the above two-parametric diagram, we have only varied
the frequency of the breathers $\wb$ and the strength of the
gain/loss $\gamma$. To illustrate how the results vary
as the final (coupling) parameter of the system varies, we
have illustrated the same features as in
Figs.~\ref{fig:softplane}-\ref{fig:examp1} for roughly half
the coupling strength in Fig.~\ref{fig:softplane2}.
We observe that the region of stability of the different
solutions (and especially of the symmetric one) has non-trivially
changed upon the considered parametric variation.
Nevertheless, sufficiently close to the linear limit
of emergence of the two solutions, the RWA remains
a reasonable description of their existence and stability, as well
as of the saddle-center bifurcation leading to their disappearance.
On the other hand, as one further deviates from this limit
towards lower frequencies, the RWA fails to capture the observed
phenomenology by deviating from the critical point for the
saddle-center bifurcation, missing the complex boundary of stability
of the symmetric branch and missing altogether the destabilization
of the anti-symmetric branch.

\begin{figure}[t]
\begin{center}
\begin{tabular}{cc}
    \includegraphics[scale=0.4]{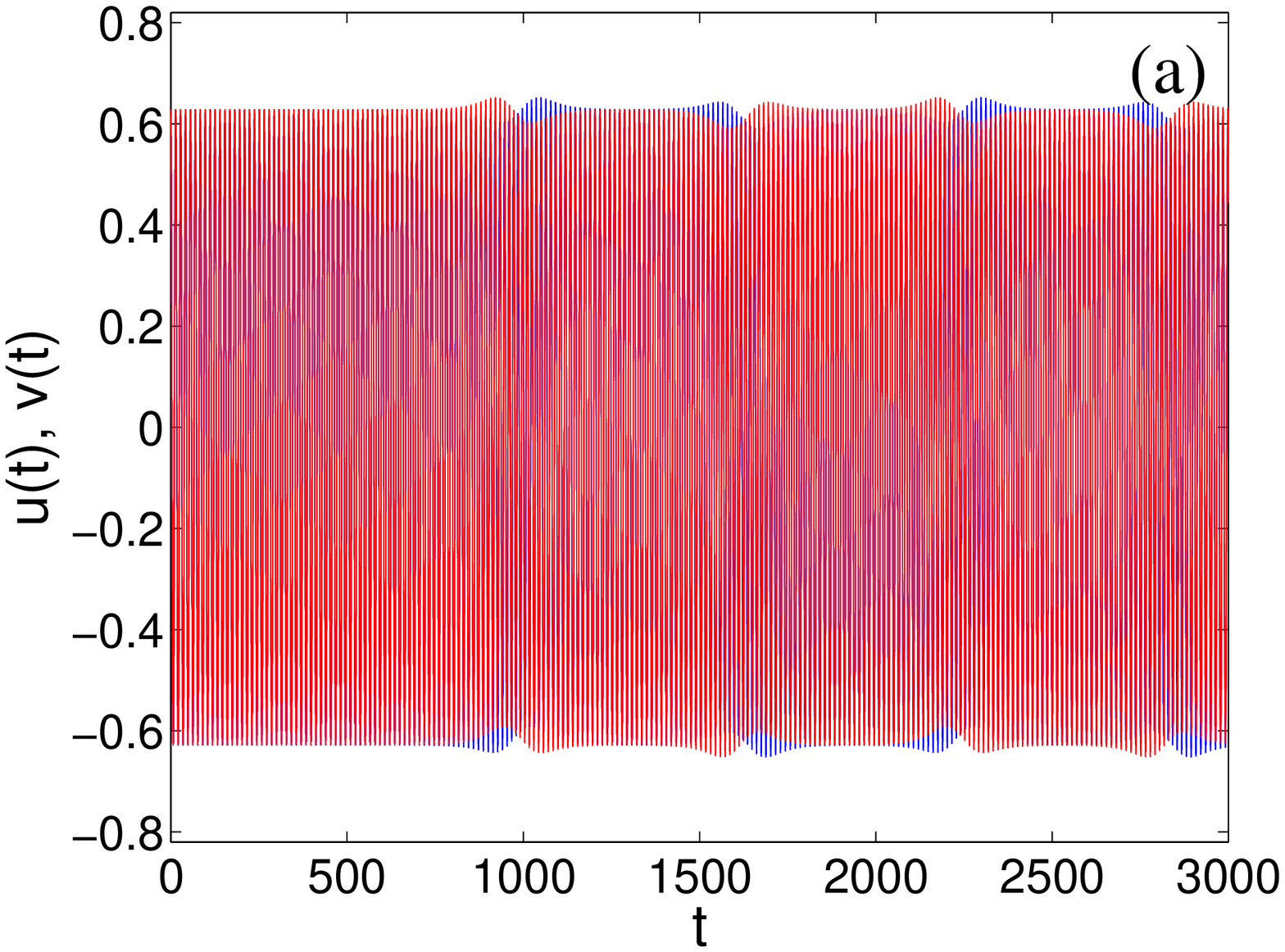} &
    \includegraphics[scale=0.4]{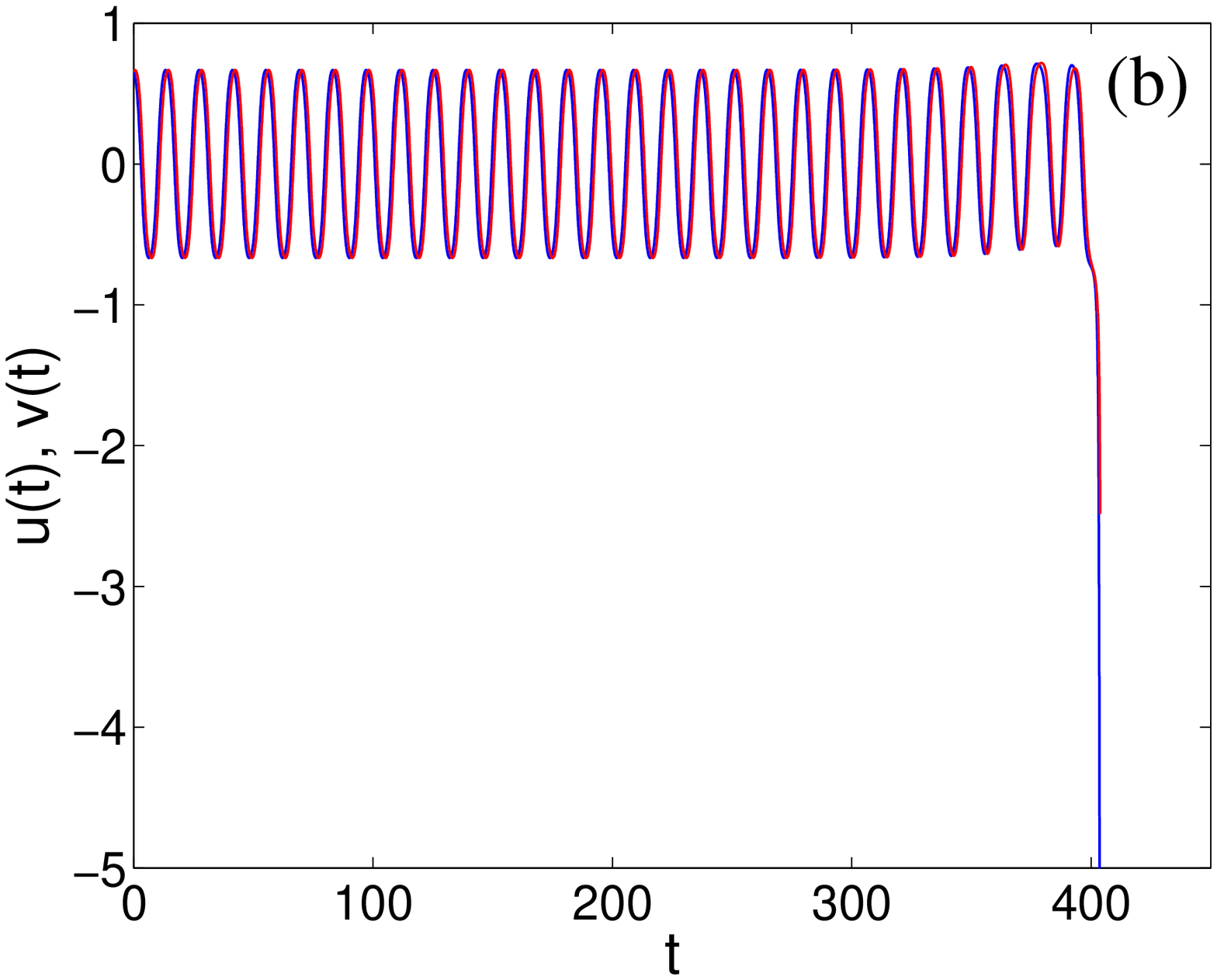} \\
    \includegraphics[scale=0.4]{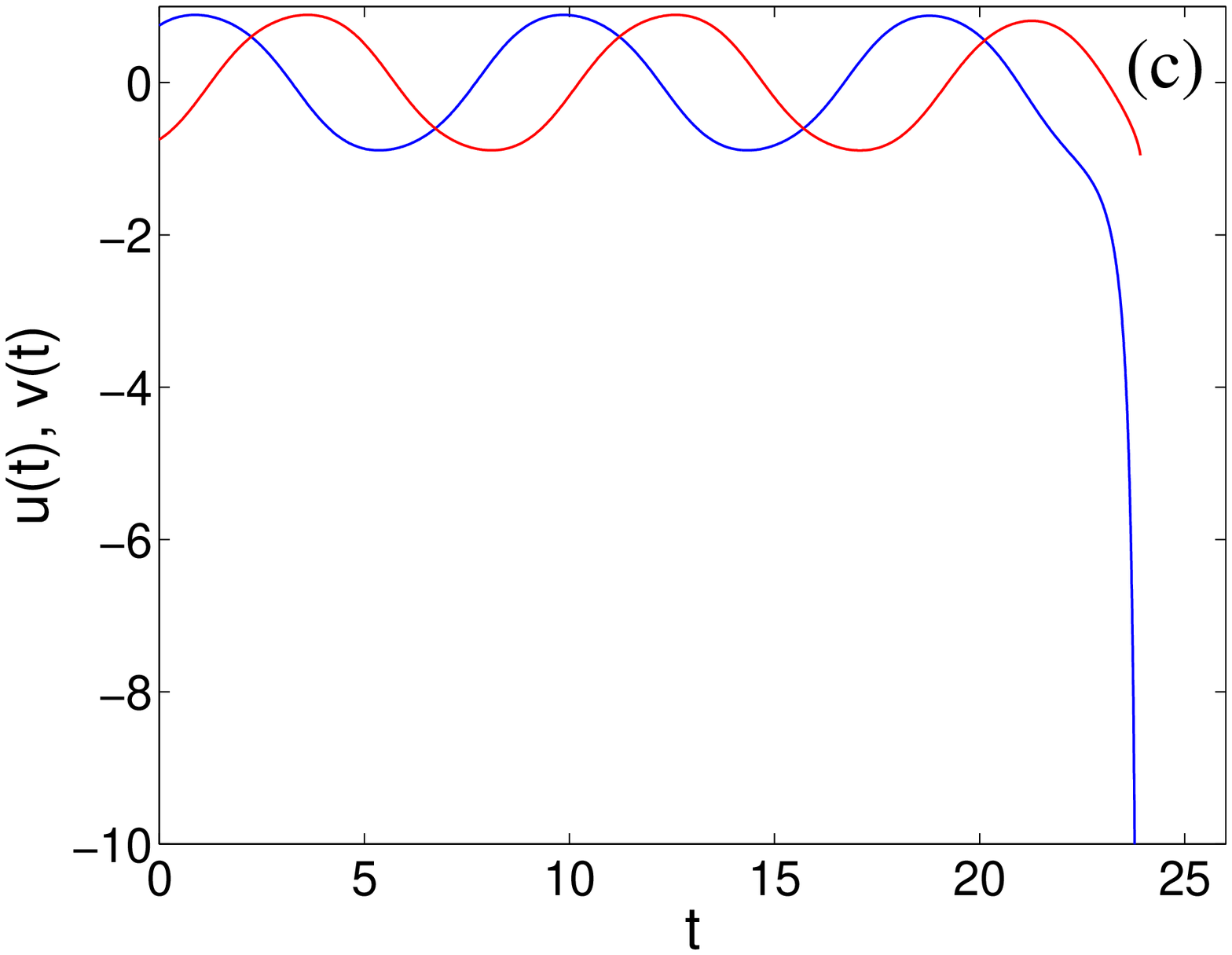} &
    \includegraphics[scale=0.4]{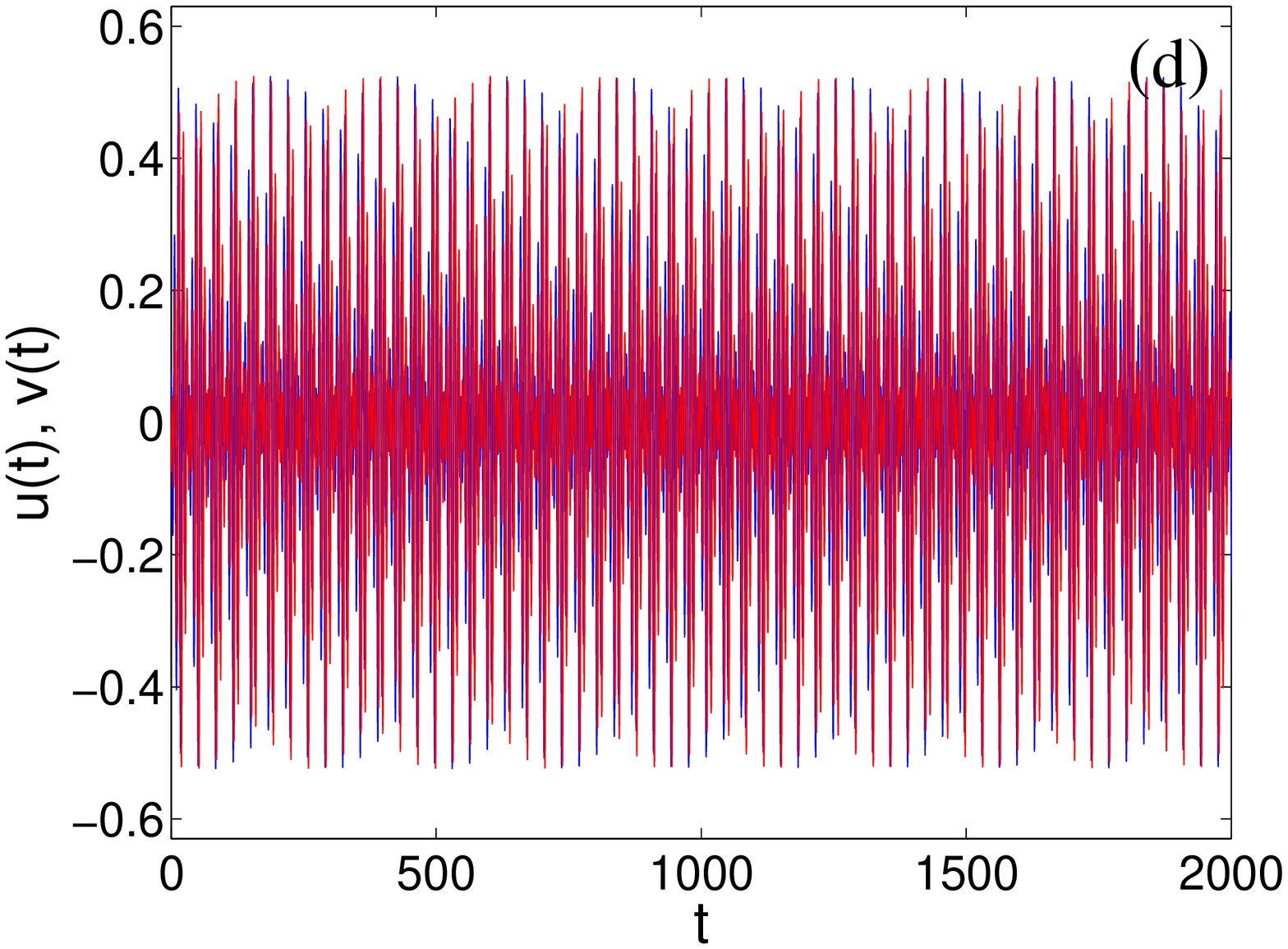} \\
\end{tabular}
\caption{(Color Online) Evolution of unstable solutions for the soft potential; the instability is driven {\em only} by the numerical truncation errors. (Top left panel) S solution with $\wb=0.45$ and $\gamma=0.01$. (Top right panel) S solution with $\wb=0.45$ and $\gamma=0.45$. (Bottom left panel) A solution with $\wb=0.70$ and $\gamma=0.66$. (Bottom right panel) The A solution at $\gamma=0.16$ and $\wb=1.2$ is taken at initial condition for $\gamma=0.51$; notice that for this value of $\wb$, the saddle-center bifurcation takes place at $\gamma_{SC}=0.168$.} \label{fig:dynamics1}
\end{center}
\end{figure}

Some examples of the evolution of unstable solutions for $s=\sqrt{15}/8$ are
shown at Fig.~\ref{fig:dynamics1}. In these cases, the perturbation
was induced solely from numerical truncation errors. In most cases, the
instabilities lead to finite time blow-up, whereas in some cases a switching
between the oscillators is observed i.e., a modulated variant of the
time-periodic solution arises as a result of the instability. However,
generally speaking, if the perturbation is above a threshold and/or the
value of the gain/loss parameter $\gamma$ is sufficiently high and/or
the solution frequency sufficiently low,
the instability manifestation will typically
result in a {\it finite time} blow-up.
This is strongly related to the escape dynamics considered in~\cite{escape}.
On the contrary, we want to highlight that this is different than the
``worst case scenario'' of the Schr{\"o}dinger dimer of the RWA.
As illustrated in~\cite{Pelinov}, in the latter at worst an exponential
(indefinite) growth of the amplitude may arise (but no finite time blow up).
Our numerical computations indicate that
the switching behavior reported above is only possible provided that the
growth rate (i.e. the FE) of the periodic solution
is small enough. For S (A) solutions, this can be achieved close to curve 4 (3). This condition is, however, not sufficient, as shown in the top right panel of Fig.~\ref{fig:dynamics1}. We have also analyzed the outcome for solutions with $\gamma>\gamma_{SC}$, taking as initial condition a solution for $\gamma<\gamma_{SC}$; in that case, we have observed that, although the generic scenario corresponds to a blow-up, isolated cases of modulated dynamics may arise when such a profile is used as initial condition for the simulation.

\subsection{Hard potential}

We now briefly complement these results with ones of the far more
accurately approximated (by the RWA) hard potential
case. This disparity in the much higher level of adequacy of the theoretical
approximation here is clearly induced by the
absence of finite-time collapse in the latter model in consonance (in this
case) with its RWA analog.

In this case, the branches exist to the {\it right} of the linear
curve (for a given $\gamma$), as is expected in the case of
a hard/defocusing nonlinearity. Curve 1 illustrates the
linear limit; once again its growing part ($\omega_1$) is associated
with the symmetric solutions, while its decreasing part ($\omega_2$)
with the ant-symmetric solutions, with their collision point
representing the linear $\cP \cT$ transition point. In fact,
from that point, emanates curve 2 which is the nonlinear $\cP \cT$ phase
transition curve i.e., the locus of points where the S and A
solutions collide and disappear for the nonlinear problem.
Notice the very good comparison of this curve with the theoretical
prediction of the RWA for $\gamma_{SC}$.
Curve 3, on the other hand, denotes the point of destabilization
of the A branch, which is in fact expected also from the RWA,
whose prediction is once again (red dashed line) in remarkable
agreement with the full numerical result. The final curve in the
graph, i.e., the green solid line of Curve 4 denotes a narrow
parametric window beyond which (or more appropriately between which
and Curve 2) even the S branch of time-periodic solutions is
destabilized. This, as indicated previously, is a feature that
is {\it not} captured by the RWA but is unique to the oscillator
system (as is correspondingly the destabilization of the A branch
in the soft potential case). We now discuss the existence
and stability of the branches in each of the regions between
the different curves.

\begin{figure}[t]
\begin{center}
    \includegraphics[scale=0.4]{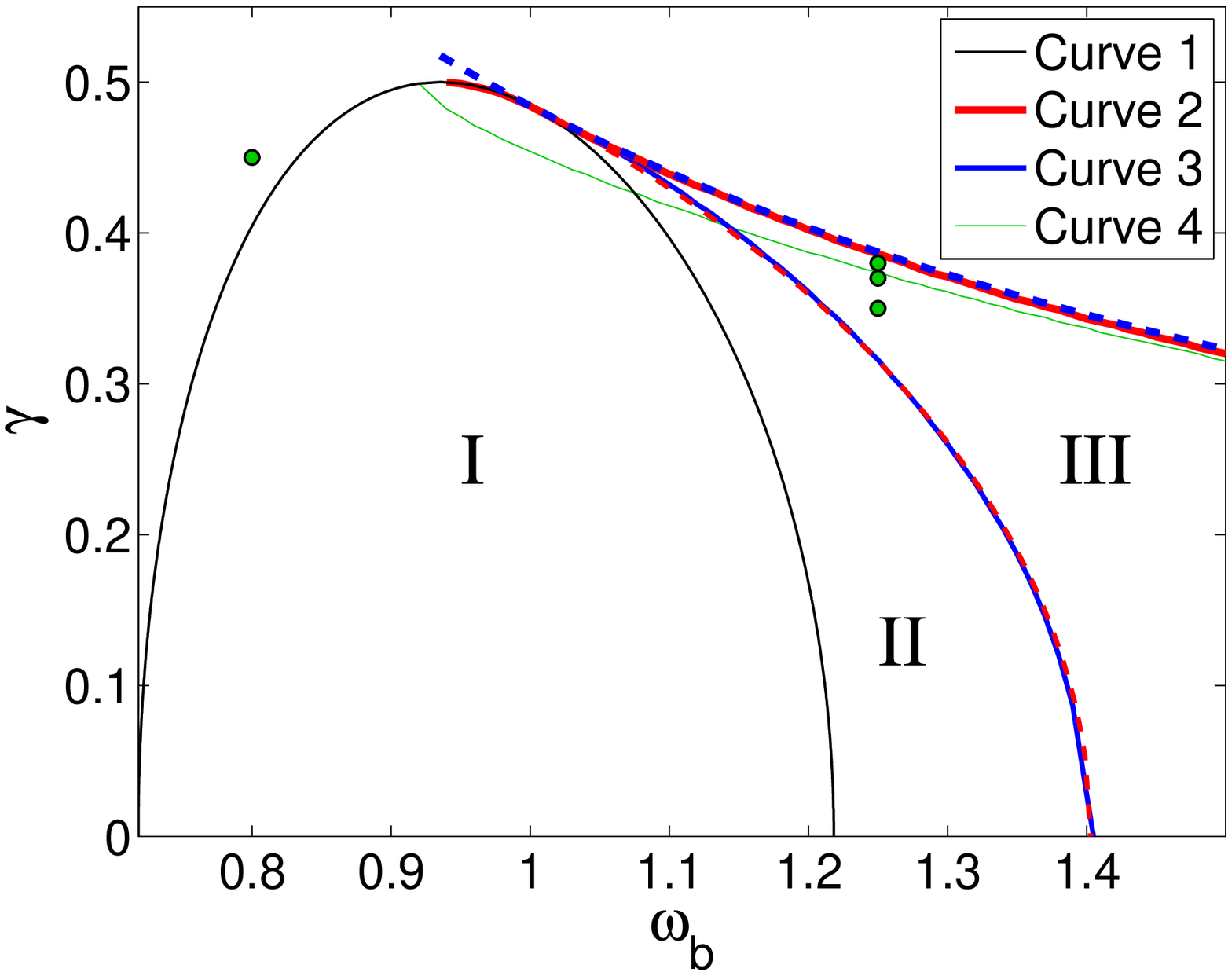}
\caption{(Color Online) Same as Fig. \ref{fig:softplane} but in the hard potential case, i.e.
when $\epsilon=-1$. The only change with respect to that figure consists of
an interchange of the color / thickness between Curves 3 and 4. Dots mark the
parameters for which simulations are performed in Fig.~\ref{fig:dynamics2}.}
\label{fig:hardplane}
\end{center}
\end{figure}

\begin{itemize}

\item Region I: Only S solutions exist (i.e., to the right of the increasing
part of Curve 1, but to the left of its decreasing part).
They bifurcate from the linear modes when
$\wb \geq \omega_1$.
The S modes are stable in this regime.

\item Region II: Solutions S and A exist and are both stable.
The A solutions have now bifurcated for $\wb > \omega_2$.

\item Region III: Solutions S are stable whereas solutions A are unstable.
Here, the symmetry breaking bifurcation destabilizing the anti-symmetric
solutions has taken place in close accord with what is expected from the
RWA.

\item Finally, there is a Region (IV) between the (green) thin solid line of
Curve 4 and the thick solid (red) line of Curve 2 denoting the narrow
parametric regime where the S branch is unstable.

\end{itemize}

\begin{figure}[t]
\begin{center}
\begin{tabular}{cc}
    \includegraphics[scale=0.4]{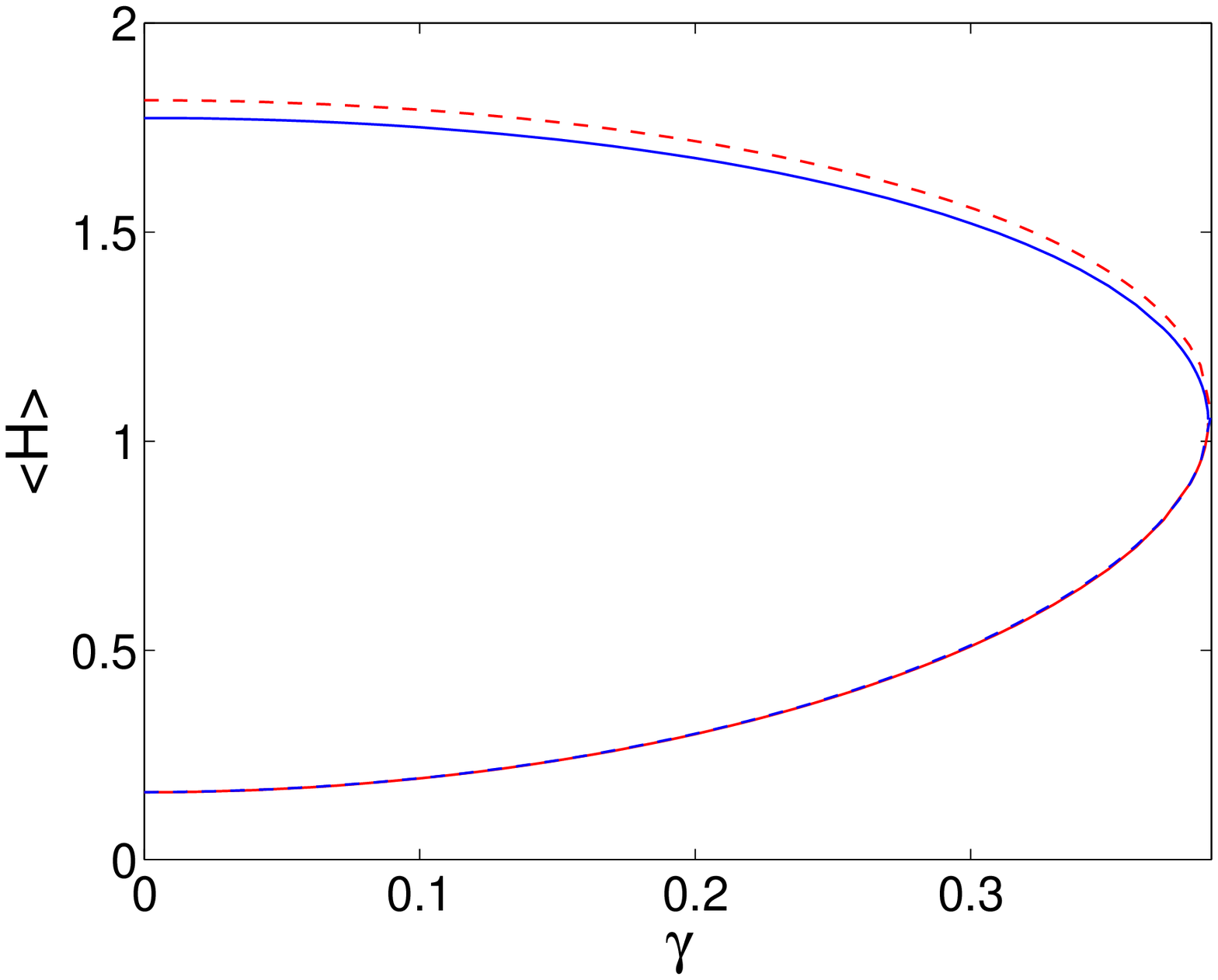} &
    \includegraphics[scale=0.4]{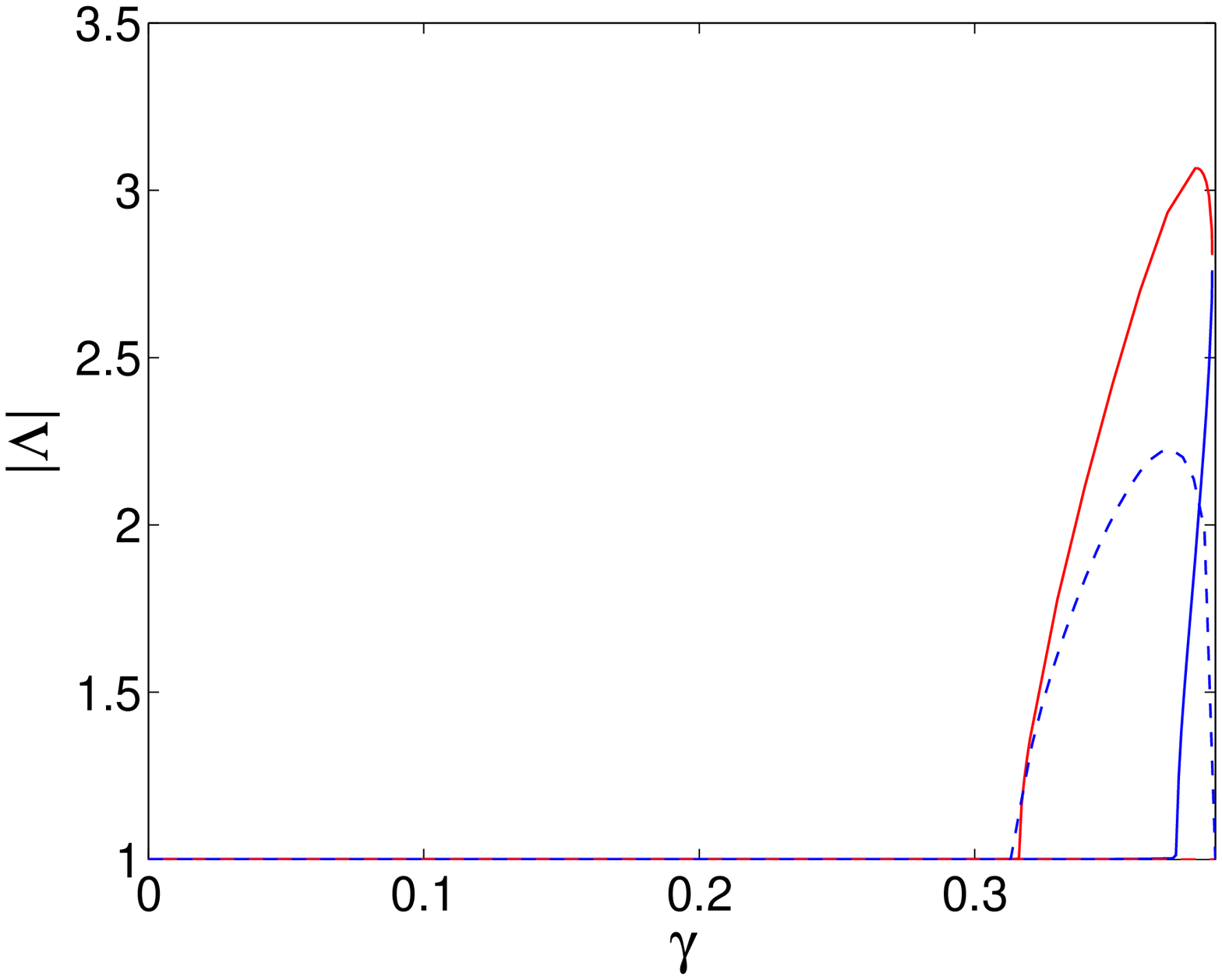} \\
    \includegraphics[scale=0.4]{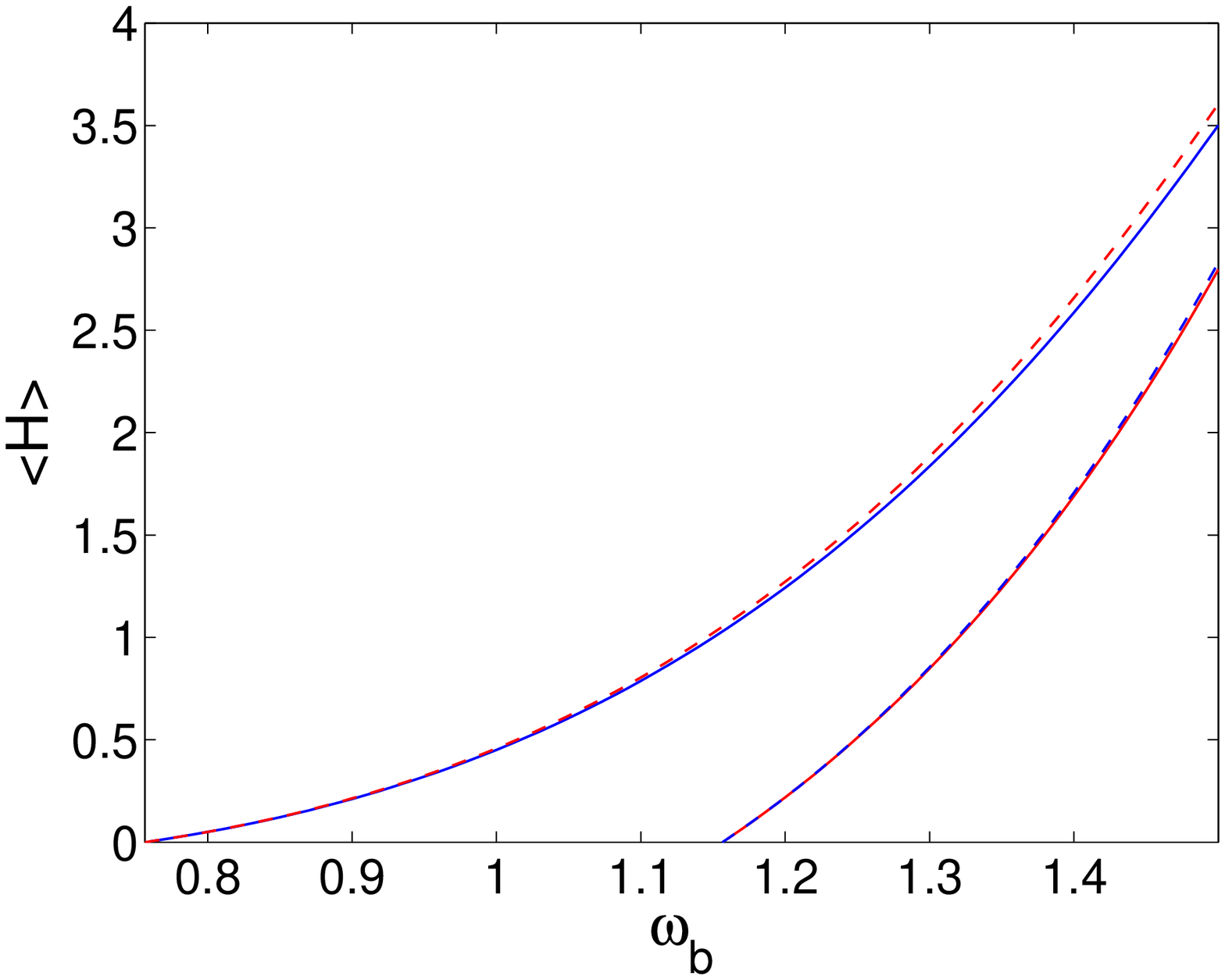} &
    \includegraphics[scale=0.4]{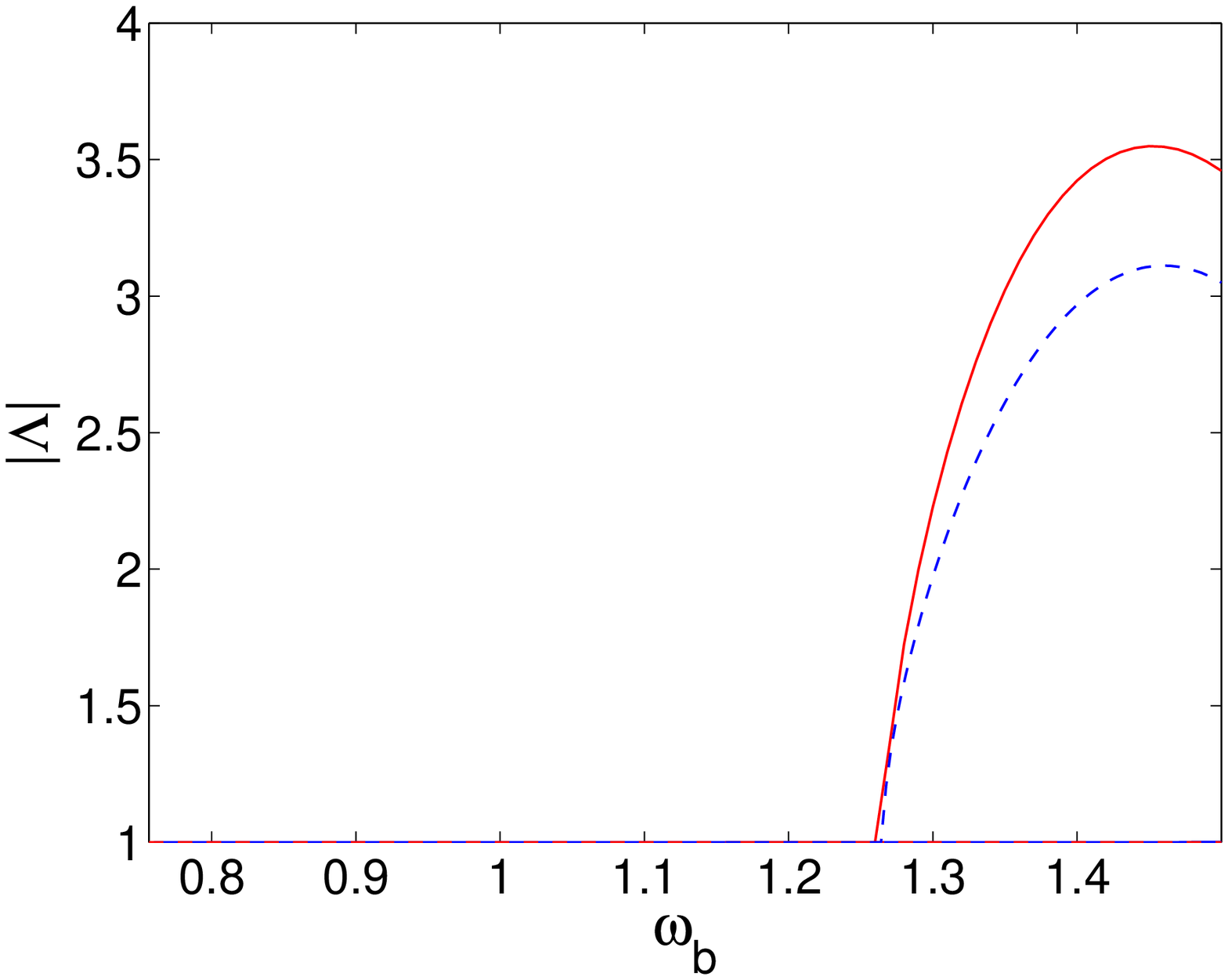} \\
\end{tabular}
\caption{(Color Online) Similar to Fig.~\ref{fig:examp1} but for $\wb=1.25$ and $\epsilon=-1$
in the top panels and for $\gamma=0.3$ and $\epsilon=-1$ in the
bottom panels. Both the existence (left) and stability (right)
diagrams of S (blue solid) and A (red solid) branches and their
rotating wave approximations (the latter with reverse colors and
dashed lines) are shown.} \label{fig:examp4}
\end{center}
\end{figure}

Similarly to Fig.~\ref{fig:examp1} in Fig.~\ref{fig:examp4}, we
offer monoparametric continuation examples (i.e., vertical or
horizontal cuts along the two parameter diagram of Fig.~\ref{fig:hardplane}.
Both along the vertical cuts of the $\gamma$ dependence for
$\wb=1.25$ in the top panels, as well as along the horizontal cuts
for $\gamma=0.3$ in the bottom panels, it can be seen that the
dashed lines of the RWA do a very reasonable (even quantitatively)
job of capturing the features of the full periodic solutions. As discussed
above, the only example where a trait is missed by the RWA is the
narrow instability interval of the symmetric branch of the blue
solid line in the top right panel of Fig.~\ref{fig:examp4}.

\begin{figure}[t]
\begin{center}
\begin{tabular}{cc}
    \includegraphics[scale=0.4]{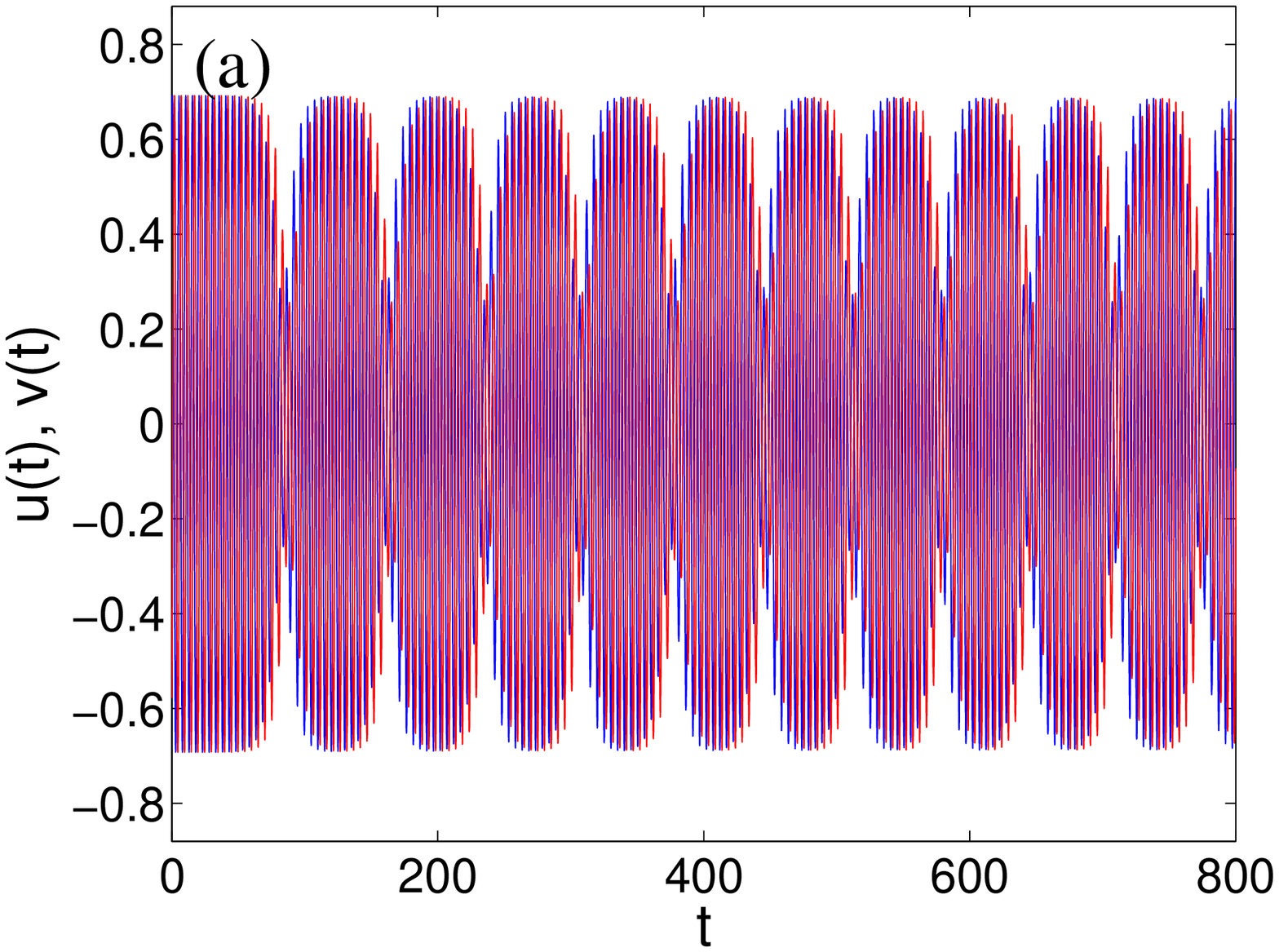} &
    \includegraphics[scale=0.4]{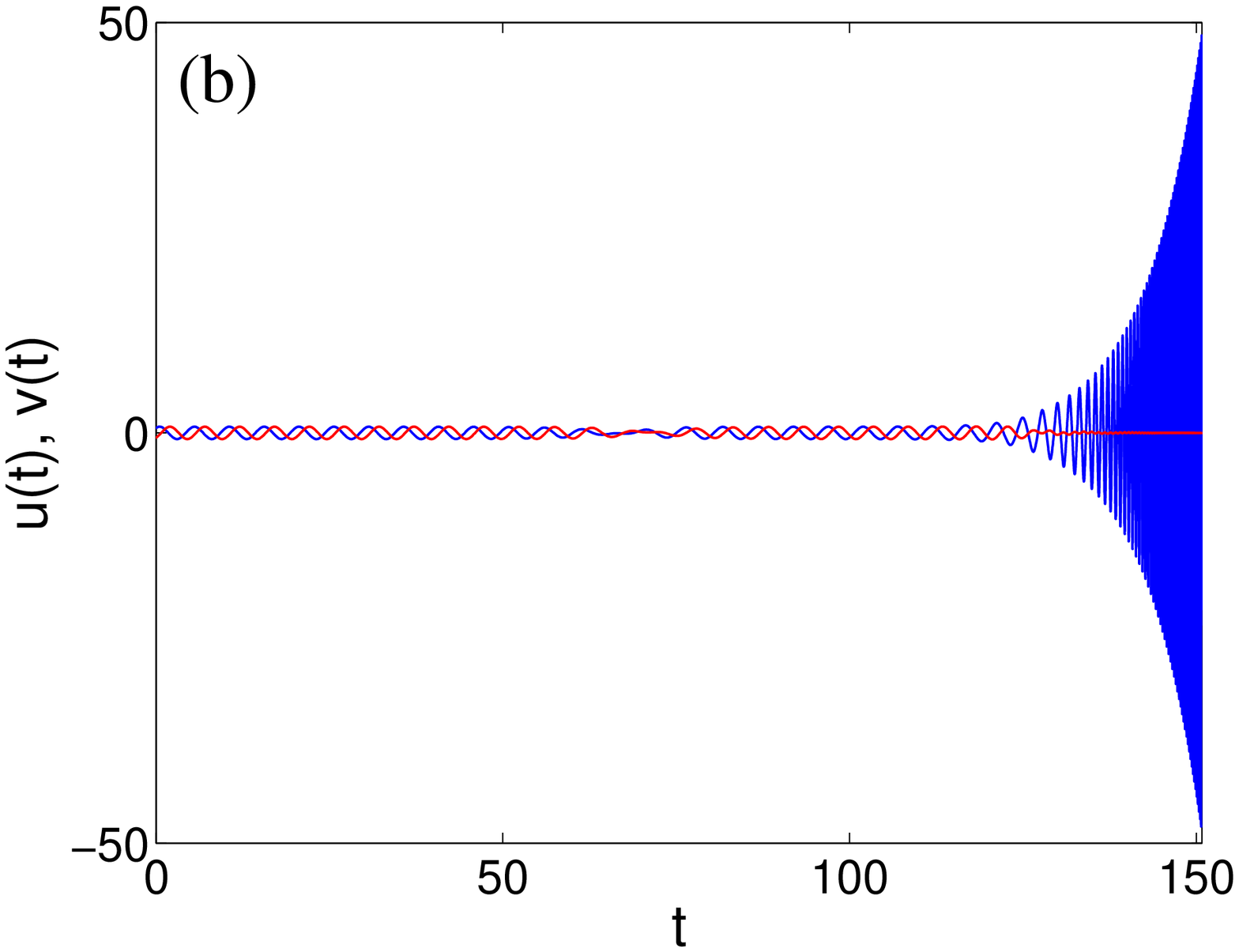} \\
    \includegraphics[scale=0.4]{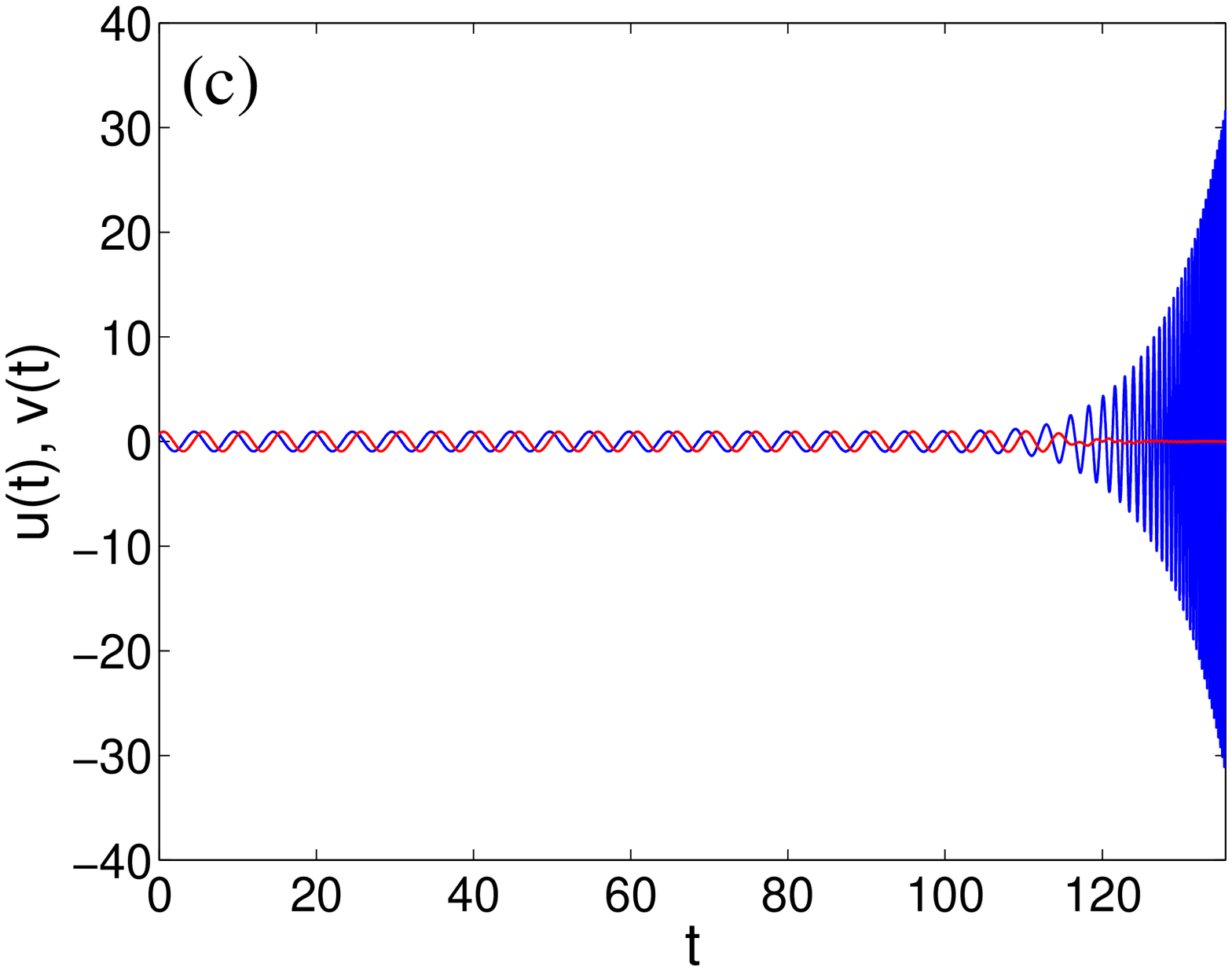} &
    \includegraphics[scale=0.4]{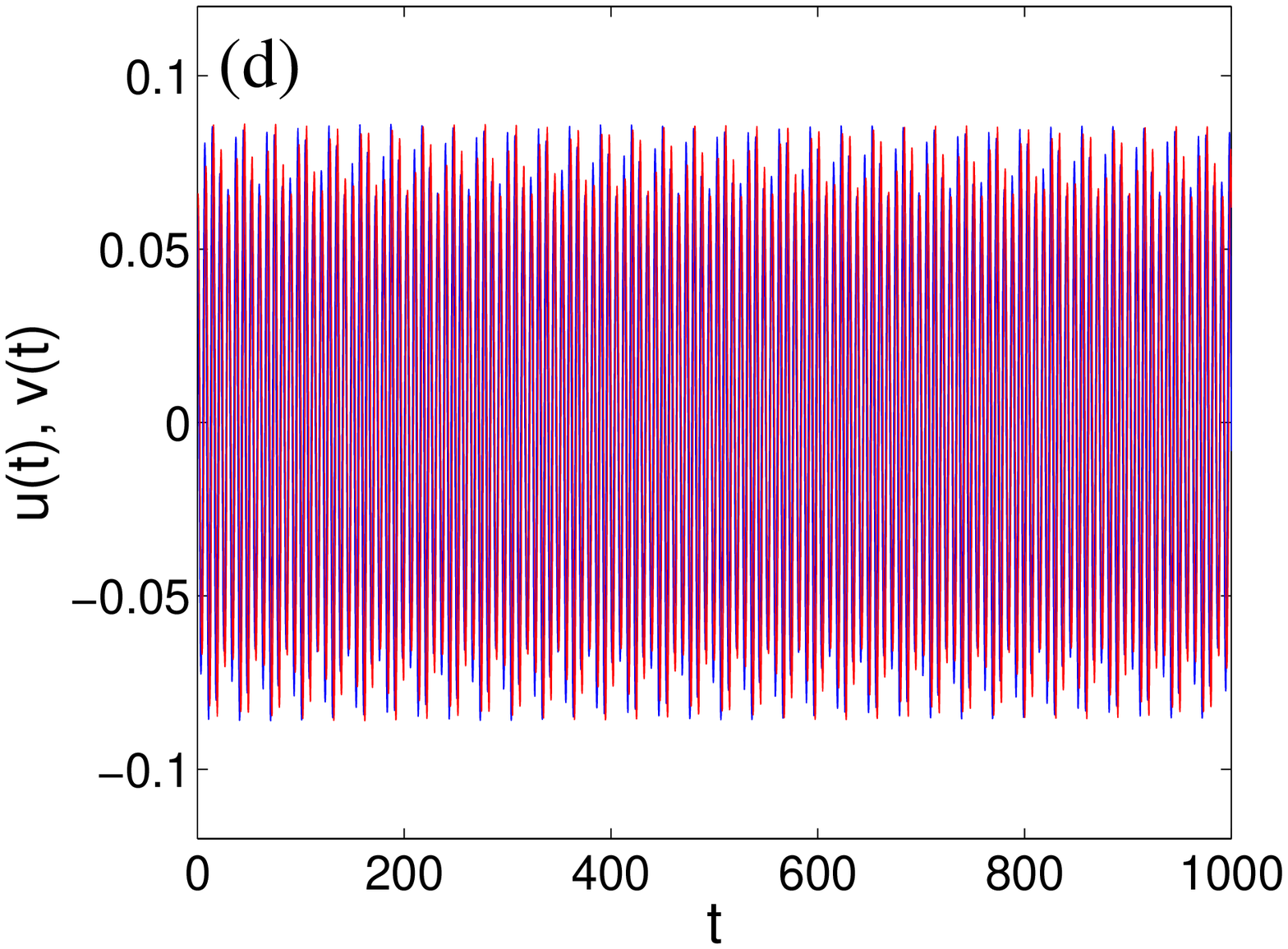} \\
\end{tabular}
\caption{(Color Online) Evolution of unstable solutions for the hard potential; the instability is driven {\em only} by the numerical truncation errors. (Top left panel) A solution with $\wb=1.25$ and $\gamma=0.35$. (Top right panel) A solution with $\wb=1.25$ and $\gamma=0.37$. (Bottom left panel) S solution with $\wb=1.25$ and $\gamma=0.38$. (Bottom right panel) The S solution at $\gamma=0.4$ and $\wb=0.8$ is taken at initial condition for $\gamma=0.45$; notice that for this value of $\gamma$, the saddle-center bifurcation takes place at $\gamma_{SC}=0.404$.} \label{fig:dynamics2}
\end{center}
\end{figure}

Two examples of the evolution of unstable solutions are shown at Fig. \ref{fig:dynamics2}. In these cases, the perturbation came only from numerical truncation errors. The most typical dynamical evolution consists of a continuous gain of the oscillator $u(t)$ which is accompanied by
a vanishing of the oscillations of $v(t)$.
However, notice that in this case and in agreement with the expectations
from the RWA and Schr{\"o}dinger $\cP \cT$-symmetric dimer, the indefinite
growth does not arise at a finite time (i.e., finite time blowup) but
instead emerges as an apparent (modulated) exponential growth on the
one node, associated with decay in the second one.
However, we should highlight an additional possibility which can
arise in the form of a quasi-periodic orbit,
if the perturbation is small enough. This dynamical behavior can turn into
the gain / vanishing one, as the size of the perturbation is increased. {Let us mention that although A solutions are mostly prone to indefinite growth, there are some cases where instability could manifest itself as switching;
at a given value of $\wb$, our simulations indicate that
switching occurs at a range of intermediate values of $\gamma$. For instance,
for $\wb=1.25$, the A solution is unstable in the range
$\gamma\in[0.316,0.386]$ and switching (i.e., modulation of the periodic orbit)
takes place for $0.35\lesssim\gamma\lesssim0.36$. S solutions offer a similar
trend.}

\section{Conclusions and Future Challenges}

In the present work, we have considered a prototypical example
of a $\cP \cT$-symmetric dimer in the context of coupled nonlinear
oscillators. We explored how the behavior of the Hamiltonian
limit of the system is modified in the presence of the
gain/loss parameter $\gamma$ which plays a significant
role in the dynamics. We were able to quantify the existence,
stability and nonlinear dynamics of the model numerically
by means of a Newton solver identifying periodic orbits,
coupled to a Floquet exponent computation, as well as
a time-stepper of the system's evolution. We also provided
a useful theoretical approximation to the relevant
features by means of a Schr{\"o}dinger dimer via the rotating
wave approximation but also delineated the limitations of such
an approach. Through the combination of these tools, we
observed how symmetric and anti-symmetric periodic orbits bifurcate
from a quantifiable linear limit, how they become unstable through
symmetry-breaking bifurcations and finally how they terminate
in a nonlinear analog of the $\cP \cT$ transition. While most
of these features could be theoretically understood by means
of our (linear and nonlinear RWA) analysis, we also revealed
a set of effects {\it particular} to the oscillator system,
such as the possibility for escape and finite time collapse
in the case of soft nonlinearity, as well as the potential
for destabilization of {\it both} branches (rather than
just the single one suggested by RWA). We also explained
the regime where the RWA was expected to be most efficient
(i.e., for parameters proximal to the bifurcation from the
linear limit) and where more significant deviations should
be expected, most notably e.g. for much smaller frequencies $\wb$ in
the soft potential case.

This work, we believe, paves the way for a number of future
considerations in the context of oscillator problems with
$\cP \cT$ symmetry. While in the Schr{\"o}dinger context,
numerous studies have addressed the large/infinite number
of nodes limit~\cite{Pelin1,Sukh,zheng}, this is far less
so the case in the context of oscillators where this
analysis is really at a nascent stage. In such lattice
contexts, it would be of particular interest to consider
genuine breather type solutions in the form of exponentially
localized in space and periodic in time orbits. Once such
structures are identified systematically in the context
of 1d settings, it would also be natural to extend consideration
to higher dimensional plaquettes~\cite{Guenter} and lattices,
where more complex patterns (including discrete vortices
among others~\cite{aubry}) are known to be possible. Such
studies are currently in progress and will be reported in
future publications.


\end{document}